\documentclass[10pt]{iopart}

\usepackage{iopams}
\usepackage{graphicx}
\usepackage{multicol}
\usepackage{psfrag}
\usepackage{amssymb}
\usepackage{amsfonts}
\usepackage{bm}
\usepackage{pstricks}
\usepackage{siunitx}
\usepackage{color}

\begin{document}


\title[Entangled photon emission from InGaN QD systems]{Indium gallium nitride quantum dots: Consequence of random alloy fluctuations for polarization entangled photon emission}

\author{Saroj Kanta Patra$^{1,2}$ \& Stefan Schulz$^{1}$}

\address{$^1$ Tyndall National Institute, University College Cork, Cork, Ireland}
\address{$^2$ Department of Electrical Engineering, University College Cork, Cork, Ireland}
\ead{sarojkanta.patra@tyndall.ie} \vspace{10pt}
\begin{indented}
    \item[]\today
\end{indented}

\begin{abstract}
We analyze the potential of the $c$-plane InGaN/GaN quantum dots for
polarization entangled photon emission by means of an atomistic
many-body framework. Special attention is paid to the impact of
random alloy fluctuations on the excitonic fine structure and the
excitonic binding energy. Our calculations show that $c$-plane
InGaN/GaN quantum dots are ideal candidates for high temperature
entangled photon emission as long as the underlying
$C_{3v}$-symmetry is preserved. However, when assuming random alloy
fluctuations in the dot, our atomistic calculations reveal that
while the large excitonic binding energies are only slightly
affected, the $C_{3v}$ symmetry is basically lost due to the alloy
fluctuations. We find that this loss in symmetry significantly
impacts the excitonic fine structure. The observed changes in fine
structure and the accompanied light polarization characteristics
have a detrimental effect for polarization entangled photon pair
emission via the biexciton-exciton cascade. Here, we also discuss
possible alternative schemes that benefit from the large excitonic
binding energies, to enable non-classical light emission from
$c$-plane InGaN/GaN quantum dots at elevated temperatures.
\end{abstract}

%
%
\ioptwocol
%

%
%
%
%
%

\section{Introduction}

\begin{figure}[b]
\centering
\includegraphics[width=0.75\columnwidth]{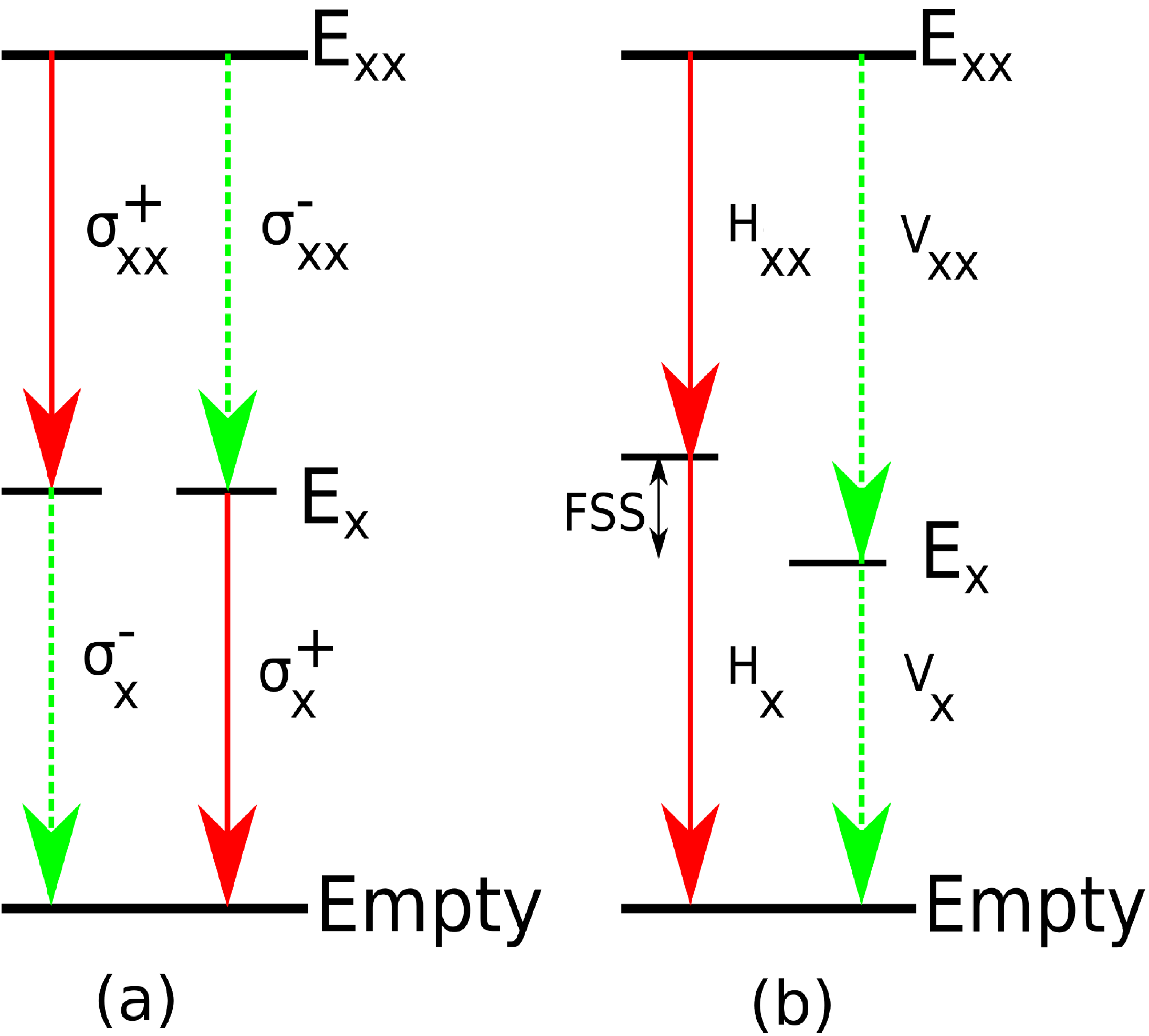}
\caption{Schematic illustration of the light polarization properties
of photons emitted from a biexciton-exciton cascade. (a) Vanishing
fine structure splitting $\Delta_\text{FSS}=0$. In this case the
photons are circularly ($\sigma_{\pm}$) polarized. (b) Non-vanishing
fine structure splitting, the emitted photons are linearly
polarized, either horizontally (H) or vertically (V) to a given
reference axis.}
    \label{fig:finestructuresplitting}
\end{figure}

Non-classical light sources capable of efficiently emitting
entangled photon pairs is one of the basic requirements to realize
secure quantum communication~\cite{GiRi2002,Kimble2008,HuRe2018}.
Several methods have been employed in the past to realize highly
efficient entangled photon sources~\cite{OrVe2017,HoAb2012}. For
instance, parametric down conversion of atomic based systems is one
of the widely used method in this regard.~\cite{GiGi2002,ZhJi2011}
However, photon generation through this scheme is probabilistic and
reduces the overall efficiency.~\cite{HuRe2018} For this reason,
semiconductor quantum dots (QDs) have been a topic of enormous
scientific research interest in recent years, since in principle
they can generate polarization entangled photon pairs
deterministically by utilizing the biexciton-exciton
cascade~\cite{AkLi2006,MuFa2009}. The underlying concept in an
\emph{ideal} structure is schematically illustrated in
Fig.~\ref{fig:finestructuresplitting} (a). Using this cascade and
having a degenerate bright excitonic (ground) state of energy $E_X$,
the photons emitted will be polarization entangled~\cite{PlTr2012}:
\begin{equation}
|\psi\rangle=\frac{1}{\sqrt{2}}\left(\left|\sigma_{XX}^{+}
\sigma_{X}^{-}\right\rangle+\left|\sigma_{XX}^{-}
\sigma_{X}^{+}\right\rangle\right)\,\, . \label{eq:entangledstate1}
\end{equation}
Here, $\sigma_{XX}^{\pm}$ ($\sigma_{X}^{\pm}$) indicates that the
emitted biexciton (exciton) photon is left/right circularly
polarized.~\cite{PlTr2012}

This scheme has been heavily studied using zincblende (ZB)
InGaAs/GaAs QDs grown along the crystallographic
[001]-direction.~\cite{BeNa2003,DuKa2011} However, early
investigations revealed that this idealized picture of degenerate
bright excitonic states is not fulfilled due to the overall $C_{2v}$
symmetry of the combined system of QD geometry and underlying ZB
crystal structure.~\cite{KaDu2010} Even in an idealized dot geometry
(e.g. square-based pyramid), the $C_{2v}$ symmetry lifts the
degeneracy of bright excitonic states, resulting in a so-called fine
structure splitting (FSS), $\Delta_{\text{FSS}}$. A schematic
illustration of this situation is depicted in
Fig.~\ref{fig:finestructuresplitting}. This small FSS  (several
tenth of $\mu$eV) leads also to a change in the light polarization
characteristics of the emitted photons; the photons are linearly
polarized, here labeled as vertical (V) and horizontal (H) to a
given reference axis. However, it should be noted that even in the
presence of a non-vanishing FSS, in principle, an ``entangled''
photon state can be realized via~\cite{OrVe2017}:
\begin{equation}
|\psi\rangle=\frac{1}{\sqrt{2}}\left(\left|H_{XX}^{}
H_{X}^{}\right\rangle+e^{\frac{i\cdot\Delta_\text{FSS}\cdot\tau}{\hbar}}\left|V_{XX}^{}
V_{X}^{}\right\rangle\right)\,\, , \label{eq:entangledstate2}
\end{equation}
where $e^{\frac{i\cdot\Delta_\text{FSS}\cdot\tau}{\hbar}}$ is a
phase factor which depends on the product of FSS and exciton
lifetime $\tau$. If $\Delta_\text{FSS}$ is smaller than the
excitonic emission linewidth or $\tau$ is very small so that product
$\frac{\Delta_\text{FSS}\cdot\tau}{\hbar}\approx0$, an entangled
photon pair can be obtained.~\cite{PlTr2012}

Thus, it is of key importance to find systems that either have
intrinsically a small FSS or in which the FSS can tuned to be close
to zero. The latter has been addressed in the literature by applying
external electric or piezoelectric fields.~\cite{PlTr2012} Overall,
given that the FSS is tightly linked to the symmetry of the combined
system of dot geometry and underlying crystal structure, in ZB QDs
growth along different crystallographic directions has been
targeted. For ZB InGaAs/GaAs dots, growth along the [111]-direction
has generated significant interest, since the underlying symmetry is
$C_{3v}$ if the QD is for instance lens-shaped or a (truncated)
pyramid with a triangular base; in a $C_{3v}$-symmetric system the
FSS vanishes.~\cite{DuKa2011, KaDu2010} For [111]-oriented
InGaAs/GaAs dots this has already been demonstrated in the
literature~\cite{JuDi2013}.

While InGaAs/GaAs dots, especially when grown along the
[111]-direction, have been widely studied, material intrinsic
properties such as the small band offsets limit their application to
low temperatures. An attractive alternative to these \emph{ZB}
arsenide-based systems are \emph{wurtzite} (WZ) $c$-plane III-N QDs,
utilizing InN, GaN, AlN and their respective alloys. Ideally, such
$c$-plane III-N dots should also exhibit a $C_{3v}$
symmetry.~\cite{BaSc2007, ScCa2011} Additionally, valence and
conduction band offsets between III-N materials are very large
($\geq$ 500 meV between the binary materials).~\cite{MoMi2011}
Moreover it has been shown that exciton binding energies are very
large ($\geq$30 meV) in III-N dots.~\cite{PeDi2005,PaSc2016} Thus,
these intrinsic properties are indicative of bringing entangled
photon emission near room temperature within reach. Non-classical
light emission via single-photon emission above 200 K has already
been demonstrated in the literature for III-N
dots.~\cite{HoCh2014,HoKa2016, WaPu2017} However, compared to
InGaAs/GaAs QDs, entangled photon emission from III-N has been
studied far less and if so mainly for GaN/AlN dots, thus for
emission in the UV spectral region.~\cite{YaLi2014} When using
InGaN/GaN dots, the additional benefit of an in principle flexible
emission wavelength engineering in the visible spectral range is
possible, while ideally keeping all other benefits ($C_{3v}$
symmetry, large band offsets, large excitonic binding energies).
However, the study of the potential of InGaN/GaN QDs for entangled
photon emission is sparse, and even when studied theoretically the
question of atomistic effects arising from alloy fluctuations in
InGaN is not addressed. However, it is important to note that the
optical properties of InGaN/GaN quantum wells are strongly affected
by alloy fluctuations.~\cite{GrSo2005,DaSc2016}

Here, we present a detailed theoretical analysis of the electronic
and optical properties of $c$-plane InGaN/GaN QDs by means of an
atomistic many-body theoretical framework. This framework allows us
to study the impact of random alloy fluctuations on both the
electronic structure and the excitonic FSS. We discuss resulting
consequences for entangled photon emission from $c$-plane InGaN/GaN
dots at elevated temperatures.

Our results show that when InGaN/GaN dots preserve a C$_{3v}$
symmetry, they are very promising candidates for producing
polarization-entangled photon pairs at high temperatures. For
instance, in an idealized truncated cone shaped InGaN/GaN dot
system, we observe a zero FSS between bright excitonic states and an
excitonic binding energy of approximately 50 meV. While when
including random alloy fluctuations in the description the excitonic
binding energies are still very large ($>$35 meV), and even when
keeping the truncated-cone shaped dot geometry, we find that the
random alloy fluctuations break the $C_{3v}$-symmetry. This symmetry
breaking significantly changes the excitonic fine structure, which
as we will show, has a detrimental effect for polarization entangled
photon emission; emitted photons are not orthogonally polarized.
Therefore, controlling the symmetry of $c$-plane InGaN/GaN QDs will
be key to use these dot systems for polarization entangled photon
emission at elevated temperatures. We will also briefly discuss how
entanglement may still be possible through alternative schemes.
Overall, with the added advantage of high exciton binding energies
and tunable emission wavelength, and when controlling the symmetry
of the system, $c$-plane InGaN/GaN dots are potential candidate for
high temperature non-classical light emission over a wide wavelength
range.

The paper is organized as follows. We start in Sec.~\ref{sec:Theory}
with a brief discussion of the theoretical framework. The full
details are given in an Appendix. Section~\ref{sec:QDgeo} presents
the QD geometry used in this study. In
Sec.~\ref{sec:SPstates_energies} we analyze the impact of alloy
fluctuations on the single particle states and energies, while in
Sec.~\ref{sec:X_FSS} the excitonic FSS, excitonic binding energies
and light polarization characteristics of the emitted photons are
presented and analyzed. In all these investigations special
attention is paid to the impact of random alloy fluctuations on the
results. Finally, we summarize our work in
Sec.~\ref{sec:Conclusion}.

\section{Theoretical framework}
\label{sec:Theory}

In this section we briefly describe the theoretical background
underlying our calculations of electronic and optical properties of
$c$-plane InGaN/GaN QDs. More details of the underlying methods can
be found the Appendix.

On the electronic structure side we apply an $sp^3$ tight-binding
(TB) model. Since our aim is to understand the impact of alloy
fluctuations in detail, we use two different frames for the TB
model, namely a fully atomistic description and a virtual crystal
approximation (VCA). In the VCA approach we neglect any alloy
fluctuations and assume that the atoms in the dot region can be
described by virtual atoms, for which the corresponding TB
parameters are obtained as average of the TB parameters for InN and
GaN. Of central importance is that the VCA description preserves the
$C_{3v}$ symmetry of the system (see discussion below on QD
geometry). For the fully atomistic case we resolve alloy
fluctuations in the system, and set the TB parameters according to
the local atomic species. Both VCA and atomistic framework account
for strain and built-in fields. In the VCA case a continuum-based
description via a surface integral method is
applied,~\cite{WiAn2005} while in the microscopic description strain
and built-in fields are determined from valence force field and
local polarization theory methods.~\cite{CaSc2013}

Since the evaluation of the excitonic properties is intrinsically a
many-body problem, we employ the configuration interaction (CI)
scheme.~\cite{BaSc2007,WiNa2006} The CI scheme allows us to account
for electron-hole exchange interaction effects, important for the
description of the excitonic FSS, $\Delta_\text{FSS}$, and takes as
input the TB single-particle energies and Coulomb matrix elements
calculated from TB wave functions. Using Fermi's golden rule, and
calculating dipole matrix elements from TB wave functions, insight
into polarization resolved excitonic emission and absorption spectra
is gained.

Overall, our chosen theoretical framework allows us to study how
deviations from an ideally $C_{3v}$ symmetric system due to alloy
fluctuations will affect both electronic and optical properties.
Additionally, equipped with information about quantities such as
exciton binding energy, FSS and light polarization characteristics,
we will discuss the potential of $c$-plane InGaN/GaN QDs for
(elevated temperature) entangled photon emission. In order to
perform these studies, the model needs as input (i) the QD size and
geometry, and (ii) for the fully atomistic description the In atom
distribution in the dot. We also note that when studying the
electronic structure and optical properties, and in particular the
FSS, of $c$-plane WZ QDs, the choice (symmetry) of the supercell is
extremely important. If the latter aspect is not carefully
considered, the FSS values predicted are potentially too large. All
these aspects will be discussed in the following section.

\section{Quantum dot geometry, simulation supercell and alloy configurations}
\label{sec:QDgeo}

\begin{figure}[t]
    \centering
    \includegraphics[width=1\columnwidth]{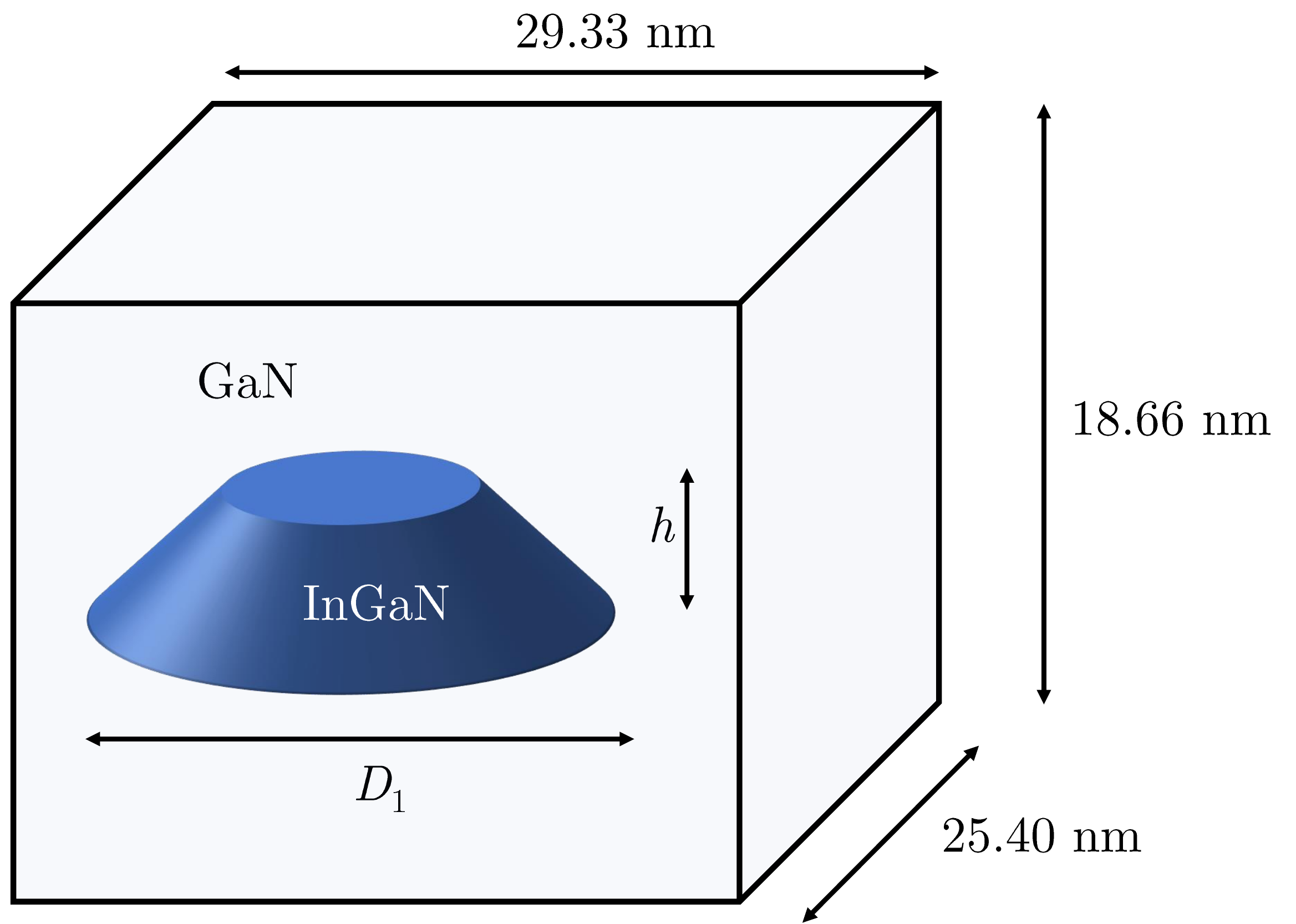}
    \caption{Schematic illustration of a truncated-cone shaped $c$-plane
        InGaN QD buried in a GaN matrix considered in this study. The
        diameter (height) of the QD is denoted by $D_1$ ($h$). Additionally,
        the supercell dimensions are given; the considered system
        contains 1,218,816 atoms} \label{fig:QDgeoch6}
\end{figure}

As mentioned above, the theoretical framework requires QD geometry
and size as an input as well as the In content. Recent structural
investigations on $c$-plane InGaN QDs,~\cite{WoNi2017} using
transmission electron microscopy (TEM), revealed truncated
cone-shaped geometries for such dots. The reported dot diameter are
in the range of $\approx$10-20 nm, with heights varying between 3-5
nm.~\cite{WoNi2017} The In content of experimentally realized
InGaN/GaN dots is typically in the 20\% to 25\%
range.~\cite{OlBr2003, JaOl2007} Building on these recent
experimental results, we have assumed a truncated-cone shaped InGaN
QD with a base diameter $D_1$= 13 nm and a height $h=3$ nm; the In
content of our dot is 20\%. The QD geometry is schematically shown
in Fig.~\ref{fig:QDgeoch6}. Previous theoretical studies on
$c$-plane InGaN QDs have also assumed similar
geometries~\cite{TsHo2010,HoKu2010}.

Having established information about the general shape, size and
(average) In content, in the fully atomistic description, also the
In atom distribution in the dot region is required. For $c$-plane
InGaN/GaN quantum well structures, experimental atom probe
tomography studies reveal that the In atom distribution reflects
that of a random alloy.~\cite{HuGr2017,BuLa2018} Therefore, in a
truncated cone-shaped area of the simulation cell, Ga atoms are
replaced randomly by In atoms.

For the VCA calculations, this microscopic feature is lost, since
within the dot all atoms are treated as virtual atoms, described by
TB parameters interpolated between InN and GaN TB parameters. The
virtual atoms are placed on an ideal WZ lattice. Therefore, due to
the dot geometry and the VCA description, the symmetry of the system
is in principle $C_{3v}$. However, care must be taken when
generating a supercell for the calculation of electronic and optical
properties.~\cite{Patra2020} Ideally, and to preserve the $C_{3v}$
symmetry, the supercell could for instance be hexagonal, so that a
rotation of $120^\circ$ around the $c$-axis/$z$-axis is a symmetry
operation of the system. Our supercell is of rectangular shape,
which in principle breaks this symmetry; thus this may lead to
artifacts in the calculated FSS. However, given that we are
interested in bound single-particle states, for which the wave
functions quickly decay away from the dot, we use large supercells
so that the boundary conditions do not impact the electronic and
optical properties of the dot; similar approaches have been
discussed by other groups.~\cite{AnORe2000} We will revisit this
aspect in more detail below. The dimensions of our supercell are
$\approx$(29.33 nm$\times$25.40 nm$\times$18.66 nm (1218816 atoms)
with periodic boundary conditions.

\begin{figure*}[t]
\centering
\includegraphics{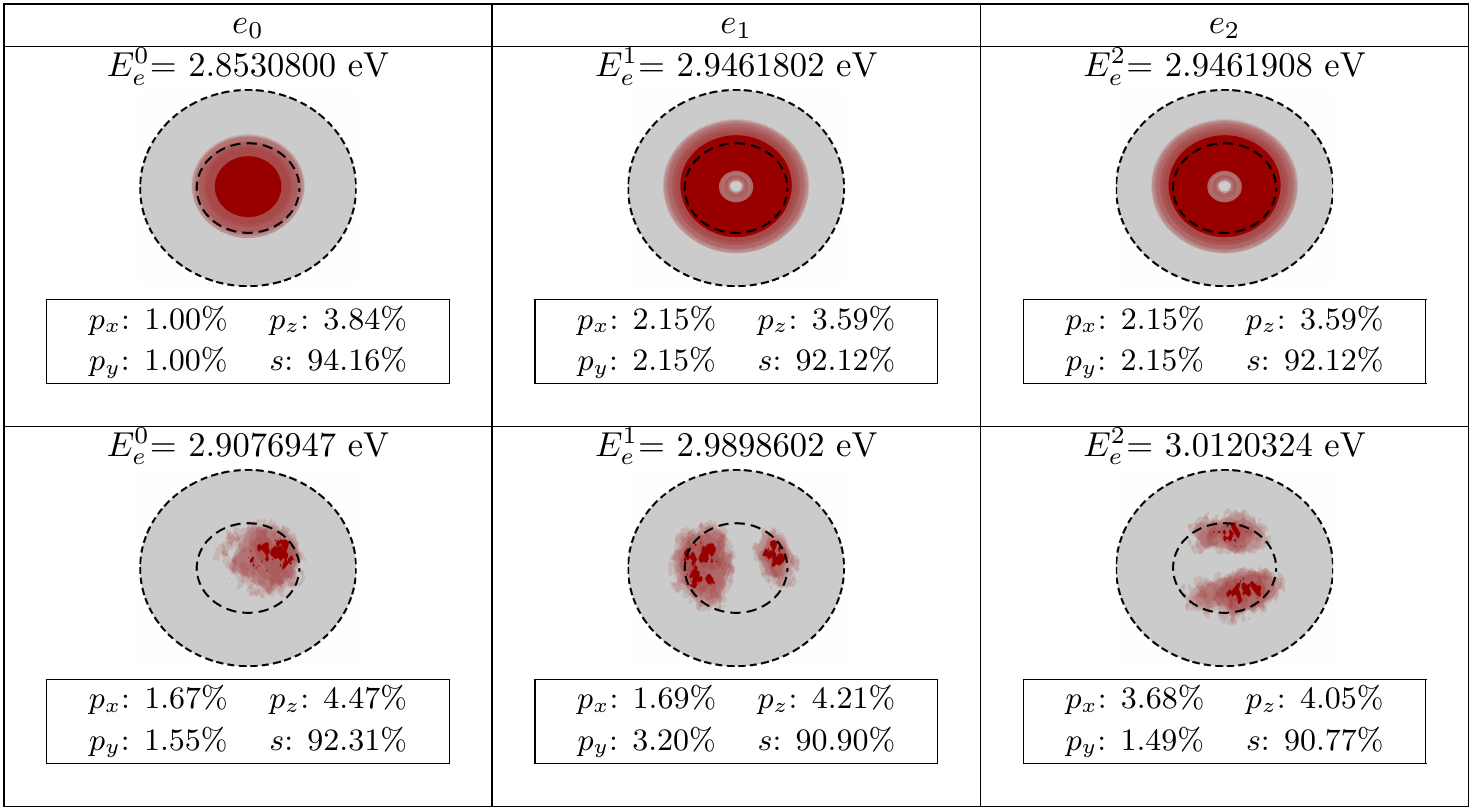}
\caption{Isosurface plots of the single-particle charge densities of
the energetically lowest three electron states ($e_0$: ground state;
$e_1$: first excited state; $e_2$: second excited state) of a
$c$-plane In$_{0.20}$Ga$_{0.80}$N/GaN QD. The light (dark) red
isosurface corresponds to 10\% (50\%) of the respective maximum
charge densities. The different sub-figures give also the
single-particles energies $E^{i}_e$ and the ($s$, $p_x$, $p_y$,
$p_z$) orbital contributions (in percent). Upper row: Results for a
virtual crystal approximation. Lower row: Same as in the upper row
but when random alloy fluctuations are described on an atomistic
level. All calculations include strain and built-in field effects as
well as spin-orbit coupling. Each single particle state is two-fold
Kramers degenerate.}
\label{fig:SP_states_electrons}
\end{figure*}

\section{Results and discussion}

In this section we present the results of our calculations. We start
with an analysis of the single-particle states and energies in
Sec.~\ref{sec:SPstates_energies}. We discuss initially the VCA
results, Sec.~\ref{sec:SPstates_energies_VCA}, followed by an
investigation of the impact of random alloy fluctuations on the
electronic structure, Sec.~\ref{sec:SPstates_energies_atomistic}. In
Sec.~\ref{sec:X_FSS}, we study the excitonic fine structure and
light polarization characteristics of the dot emission using the
VCA, Sec.~\ref{sec:X_FSS_VCA}. Subsequently, the impact of alloy
fluctuations on these quantities is analyzed,
Sec.~\ref{sec:X_FSS_Atom}. These studies are accompanied by group
theoretical arguments shedding further light onto the determined
results.

\subsection{Single particle states and energies}
\label{sec:SPstates_energies}

We start the discussion with the VCA results of the single-particle
electron and hole states. Using group theory, we also make general
statements about the electronic structure in terms of (expected)
degeneracies of energy levels and classify the different
single-particle states according to irreducible representation
(IRRs). The latter is useful for understanding and explaining the
calculated exciton fine structure. In a second step we study the
impact of the alloy fluctuations on the results.

\subsubsection{Virtual crystal approximation}
\label{sec:SPstates_energies_VCA}

\begin{figure*}[t]
\centering
\includegraphics{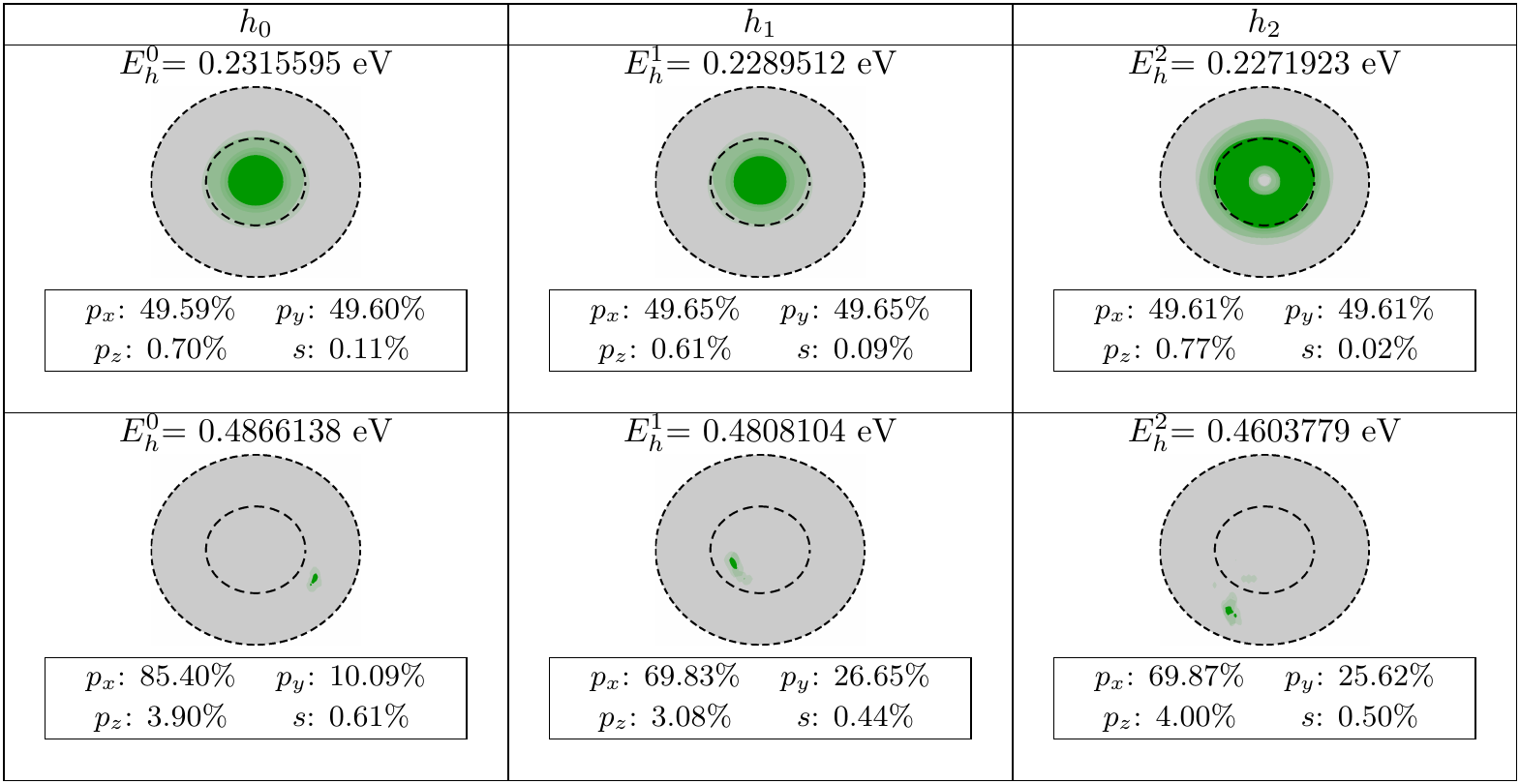}
\caption{Isosurface plots of the single-particle charge densities of
the energetically lowest three hole states ($h_0$: ground state;
$h_1$: first excited state; $h_2$: second excited state) of a
$c$-plane In$_{0.20}$Ga$_{0.80}$N/GaN QD. The light (dark) green
isosurface corresponds to 10\% (50\%) of the respective maximum
charge densities. The different sub-figures give also the
single-particles energies $E^{i}_h$ and the ($s$, $p_x$, $p_y$,
$p_z$) orbital contributions (in percent). Upper row: Results for a
virtual crystal approximation. Lower row: Same as in the upper row
but when random alloy fluctuations are described on an atomistic
level. All calculations include strain and built-in field effects as
well as spin-orbit coupling. Each single particle state is two-fold
Kramers degenerate.} \label{fig:SP_states_holes}
\end{figure*}

The upper row of Fig.~\ref{fig:SP_states_electrons} shows
isosurfaces of the probability densities $|\phi(\textbf{r})|^2$ of
the electron ground state, $e_0$, and the first two excited
single-particle states $e_1$ and $e_2$. The light and dark
isosurfaces correspond to 10\% and 50\% of the respective maximum
values. The data are displayed for a top-view/along the $c$-axis.
The gray shaded area indicates the dot geometry and the dashed lines
give the dot barrier interface. The data for the hole ground state,
$h_0$, and the first two excited states, $h_1$ and $h_2$, are
depicted in the upper row of Fig.~\ref{fig:SP_states_holes}. The
corresponding single-particle energies, also given in the respective
figures, for electrons and holes are denoted by $E^i_{e/h}$.
Additionally, the figures display the orbital contributions for each
TB single-particle state in a table below the charge densities. The
calculations include, strain and built-in field effects as well as
spin-orbit coupling (SOC) effects; each state is thus two-fold
Kramers degenerate.~\cite{ScSc2008}

Turning in a first step to the orbital character of the different
single particle states, we observe that all electron states,
cf.~Fig.~\ref{fig:SP_states_electrons}, are dominated by $s$-orbital
($\geq$ 94\%) contributions, consistent for instance with previous
$\mathbf{k}\cdot\mathbf{p}$ results.~\cite{ToVu2009, DuKa2011} For
the holes, cf.~Fig.~\ref{fig:SP_states_holes}, mainly $p_x$- and
$p_y$-orbitals contribute to the formation of their single-particle
states. The absence of a significant $p_z$ contribution is explained
by the following. First, the crystal-field splitting in a WZ crystal
breaks the symmetry between $|p_x\rangle$, $|p_y\rangle$ and
$|p_z\rangle$-like basis states. In WZ InN and GaN, the
$|p_z\rangle$-like state is shifted to a lower energy when compared
to $|p_x\rangle$ and $|p_y\rangle$-like states. Furthermore, when
looking at the effective masses of the corresponding bands in a bulk
system, the crystal field split-off band, mainly formed by
$|p_z\rangle$-like states in the absence of SOC, exhibits a lower
effective mass when compared to the topmost $|p_x\rangle$- and
$|p_y\rangle$-like bulk bands along the WZ $c$-axis. Thus, in case
of strong quantum confinement along this axis, as in the case of our
$3$ nm high $c$-plane In$_{0.2}$Ga$_{0.8}$N/GaN QD, contributions
from the $|p_z\rangle$-like basis states are suppressed in the
energetically lowest lying hole states. This explains our observed
orbital contributions for the hole states. Furthermore, given the
symmetry of the dot geometry and the WZ $c$-plane, the calculated
hole states have almost identical $p_x$ and $p_y$ orbital
contributions. In general, the observed symmetries of the wave
functions as well as the orbital contributions are consistent with
previous continuum-based calculations,~\cite{ToVu2009} again showing
that our theoretical framework captures accurately the
$C_{3v}$-symmetry of the system.

In addition to the orbital character of the different states, the
single-particle states depicted in
Fig.~\ref{fig:SP_states_electrons} and~\ref{fig:SP_states_holes} can
also be classified according to how they transform under the
symmetry operations of the point group $C_{3v}$.~\cite{BaSc2007}
Classifying the states according to the IRRs of the corresponding
point group is very helpful for analyzing the excitonic fine
structure. It has been shown and discussed by Baer~\emph{et
al.}~\cite{BaSc2007} that the electron ground state is invariant
under a rotation by $\frac{2\pi}{3}$ around the WZ $c$-axis, whereas
the states $e_1$ and $e_2$ transform as $x$- and $y$-like states
under the symmetry operations of the $C_{3v}$ group. Therefore, the
electron ground state can be denoted as an $s$-like state, while the
excited states $e_1$ and $e_2$ are $p$-like states ($p_\pm$-like
states).~\cite{Schulz2007} Using the same analysis for the hole
states, ground and first excited hole state, $h_0$ and $h_1$, can be
classified as $p$-like states, while $h_2$ is
$s$-like.~\cite{BaSc2007,Schulz2007}

We note that our calculations show only twofold (Kramer's)
degenerate states for electrons \emph{and} holes. As discussed
already in Ref.~\cite{Schulz2007}, the $C_{3v}$ \emph{single} group
(no SOC) contains two dimensional IRRs. Therefore, in such a case
and when neglecting spin degeneracies, two-fold degenerate states in
the energy spectrum (four-fold when including spin) can be expected
and are also found in previous studies for instance for the first
two excited electron states.~\cite{BaSc2005,ScSc2008} We stress that
this is in contrast to [001]-oriented InGaAs/GaAs
systems,~\cite{WiWa2000} and that in an WZ QD system that preserves
the $C_{3v}$ symmetry, strain and built-in fields do \emph{not} lift
degeneracies of the electron $p$-states.~\cite{Patra2020} This
aspect can for instance be used to test the impact of boundary
conditions on the electronic structure and that these do not lead to
artifacts in the calculations. We point out that our calculations in
the absence of SOC effects (not shown, but discussed in detail in
Ref.~\cite{Patra2020}) confirm a vanishing electron $p$-state
splitting. When including SOC effects, the symmetry of the system is
described by the \emph{double} group $\bar{C}_{3v}$. This group
contains only two-dimensional IRRs. Thus only two-fold degenerate
states are expected and the degeneracy stems from time-reversal
symmetry (Kramer's degeneracy). This behavior is indeed found in our
VCA calculations, as shown in Figs.~\ref{fig:SP_states_electrons}
and~\ref{fig:SP_states_holes}. For the hole $p$-state splitting,
$\Delta E_{h,p}=E^{0}_{h}-E^{1}_{h}$, we find a value of $\Delta
E_{h,p}=2.6$ meV, cf. Fig.~\ref{fig:SP_states_holes}, while for the
electrons the $p$-state splitting $\Delta
E_{e,p}=E^{2}_{e}-E^{1}_{e}$ is very small with a value of 10.6
$\mu$eV, cf. Fig.~\ref{fig:SP_states_electrons}. Again, in the
absence of SOC, $\Delta E_{e,p}=\Delta E_{e,p}=0$ (not shown). We
attribute the small splitting $\Delta E_{e,p}$ between the electron
$p$-states, when compared to the hole state splitting $\Delta
E_{h,p}$, to the general very small SOC energy in InN and GaN and to
the relatively weak conduction band-valence band coupling due to the
large band gap.~\cite{VuMe2003}

Overall, several important aspects should be noted. Firstly, in
[001]-oriented InGaAs/GaAs QDs, the lifting of the degeneracies of
the electron states is often used as a first indicator for a
non-vanishing FSS.~\cite{ScWi2007,KaDu2010} The fact that for instance the
electron $p$-state splitting is absent in the absence of the SOC and
only very small when including SOC, indicates already that a
different excitonic fine structure can be expected in $c$-plane
InGaN/GaN QDs. Thus, results/knowledge from [001]-oriented
InGaAs/GaAs QDs cannot be directly carried over to explain the
electronic structure or optical properties of $c$-plane InGaN/GaN
dots, as sometimes found in the literature.~\cite{NiAh2020}
Secondly, given that the wave functions and the degeneracies of the
different levels are consistent with group theoretical arguments as
it has to be the case, no artifacts in the electronic structure
calculations are introduced by the cubic supercell boundaries. This
is very important since this ensures also that the calculation of
the (small) excitonic FSS should not be affected by the boundary
conditions.

In the next step, we address the question how properties change when
treating the InGaN/GaN dot, with nominally the same In content, in a
fully atomistic framework, thus accounting for alloy fluctuations on
a microscopic level. The results of this analysis are presented in
the following section.

\subsubsection{Random alloy fluctuations}
\label{sec:SPstates_energies_atomistic}

In this section we present the single-particle states and energies
of the $c$-plane In$_{0.2}$Ga$_{0.8}$N/GaN QD when accounting for
random alloy fluctuations within the dot. In order to study the
impact of the alloy microstructure on the results, the calculations
have been performed for five different random alloy configurations.

The lower row of Fig.~\ref{fig:SP_states_electrons}
and~\ref{fig:SP_states_holes} depict a top-view of the isosurfaces
of the ground and the first two excited electron and hole state
charge densities in the presence of random alloy fluctuations.
Following the VCA description from above, the respective energies
and orbital contributions are also given in the figures. The charge
densities are displayed for an arbitrary chosen configuration, here
configuration 3 (Config.-3). The light and dark isosurfaces
correspond to 10\% and 50\% of the respective maximum values, as in
the VCA case.

Comparing the here presented states with the corresponding VCA
results (upper row), for both electrons and holes, the charge
densities are clearly deformed due to the presence of alloy
fluctuations in the dot. Even though the electron charge densities
are affected by the local fluctuations, in terms of their nodal
structure these densities still, especially for the excited states
$e_1$ and $e_2$, resemble to a first approximation $p_x$- or
$p_y$-like states. However, for the holes such a classification,
based on the nodal structure, is basically impossible. Also, due to
higher effective mass of holes when compared to electrons, the holes
localize in a smaller region as compared to electrons. Therefore,
while the localization characteristics of the electrons, at least to
a  first approximation are determined by the QD geometry and size,
for the holes alloy fluctuations within the dot lead to ``extra''
localization centers. In general, our results indicate that although
random alloy fluctuations perturb the electron wave functions, they
have a more dramatic effect on the hole wave functions. We note that
similar effects have been observed by different groups in quantum
wells.~\cite{ScCa2015,AuPe2016,JoTe2017}

Overall, given the strong perturbation of the charge densities by
the alloy fluctuations, the symmetry of the system is no longer
$C_{3v}$ but rather $C_1$.~\cite{AlHe1994} The latter symmetry means
that only a rotation of 360$^\circ$ around the WZ $c$-axis is a
symmetry operation of the system. The $C_1$ single-point group
contains only the IRR $A$ and the double group IRR
$A_{1/2}$~\cite{AlHe1994}. Both of these IRRs are 1-D; from this
analysis we can already expect a lifting of any degeneracy in the
energy spectrum of our dot system in the presence of alloy
fluctuations. We will come back to this and the consequences this
has for the excitonic fine structure in the following section.

We note also that the hole states do not necessarily localize
directly underneath the electron wave functions. This will obviously
affect the electron-hole wave function overlap and therefore further
increase the radiative lifetime $\tau$ in the system as discussed in
InGaN/GaN quantum wells already.~\cite{TaCa2016,TaDa2020} This may
also have an impact on the entangled photon emission from $c$-plane
InGaN/GaN QDs, given that for the entangled state described in
Eq.~(\ref{eq:entangledstate2}), the product of radiative lifetime
$\tau$ and FSS $\Delta_\text{FSS}$ needs to be sufficiently small.
We will revisit this question when discussing the exciton fine
structure and the light polarization characteristics of the emitted
photons in more detail.

In a final step, we discuss the orbital character of electron and
hole states of the alloy configuration (Config.-3) depicted in
Figs.~\ref{fig:SP_states_electrons} and~\ref{fig:SP_states_holes}.
The electron wave functions mainly consist of $s$-orbital
contribution with much weaker contributions from $p_x$-, $p_y$- and
$p_z$-orbitals. We observe this for all five different random alloy
configurations (not shown). This dominant $s$-orbital contribution
in the three energetically lowest electron states is also found in
the VCA results. Turning to the hole states,
Fig.~\ref{fig:SP_states_holes} reveals for the here investigated
alloy configuration, that the $p_x$-orbital contribution dominates
these states. For other configurations $p_y$-orbital contribution
dominates (not shown). This means that in contrast to the VCA-case,
where equal contributions of $p_x$ and $p_y$-orbitals are observed
in the first three bound states, alloy fluctuations break this
symmetry.

\begin{figure}[t]
\centering
\includegraphics[width=1\columnwidth]{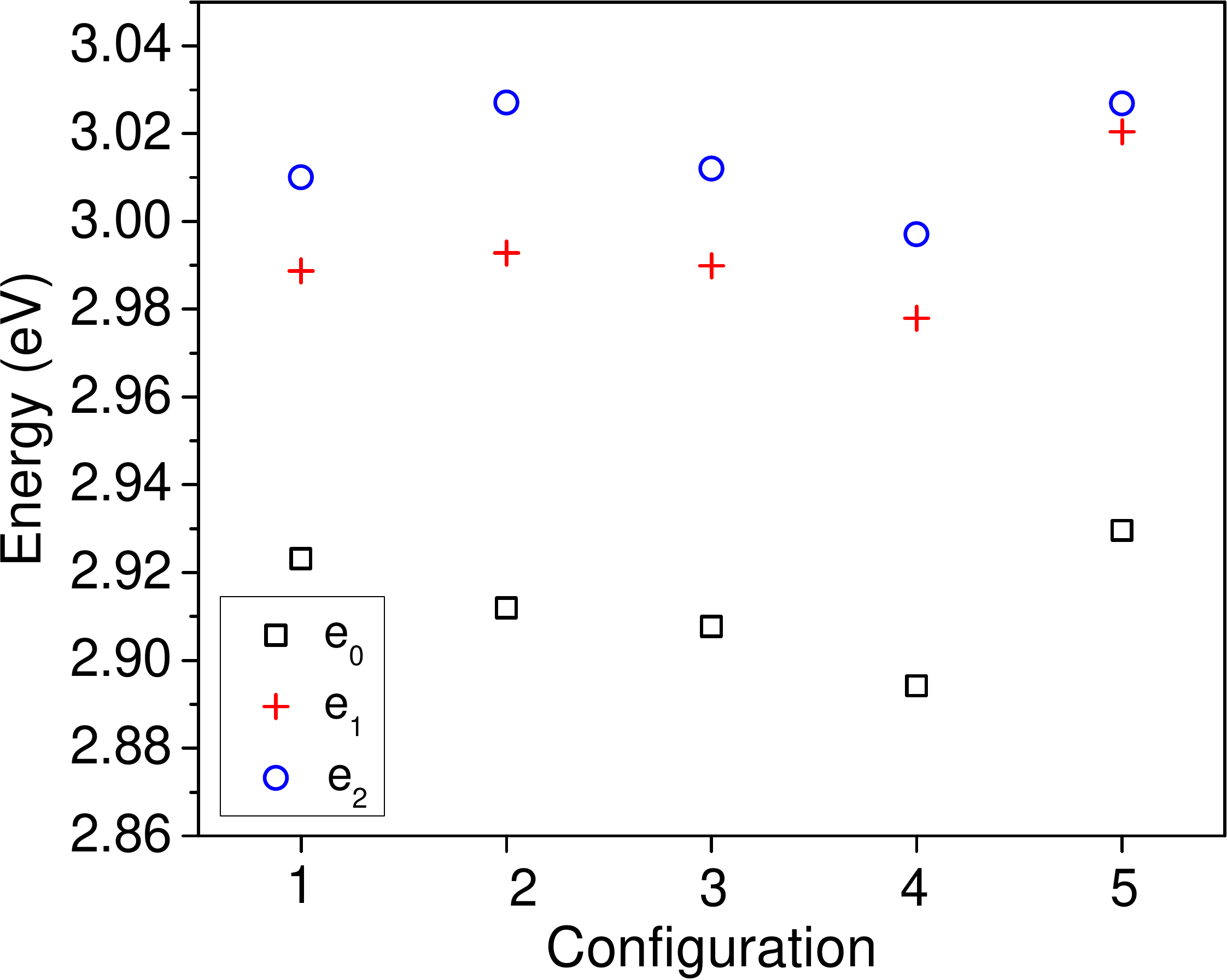}
\caption{Energies of electron ground, $e_0$, first, $e_1$, and
second, $e_2$, excited state as a function of the microscopic alloy
configurations number.}
\label{fig:randomelch6}
\end{figure}

Two consequences can now be expected from the observation that alloy
fluctuations (i) significantly affect the carrier localization
within the dot and (ii) break the symmetry between $p_x$ and
$p_y$-orbitals. Turning to (ii) first, it is important to note that
the orbital character determines to a large extent the light
polarization characteristics of the emitted photons. Thus
significant differences between the VCA and the random alloy case
can be expected in terms of the light polarization characteristics
of emitted photons. We will come back to this aspect when discussing
the calculated optical spectra. Secondly, given the strong impact of
the alloy fluctuations on the charge densities, it is also expected
that the single-particle energies are strongly dependent on the
alloy microstructure. For alloy configuration 3, the electron
$E^{i}_{e}$ and hole $E^{i}_{h}$ energies for ground and excited
states are given in Fig.~\ref{fig:SP_states_electrons} and
Fig.~\ref{fig:SP_states_holes}, respectively. Turning to the
electron energies first, we find a very large energetic separation
between states $e_1$ and $e_2$ ($p$-states) of $\Delta
E^\text{p,RA}_{e}\approx22$ meV. We note that in the VCA
calculation, the electron $p$-state splitting was very small,
$\Delta E^\text{p,VCA}_{e}=10.6$ $\mu$eV and originated from SOC
effects. As discussed already above, in InGaAs/GaAs QDs, the
electron $p$-state splitting has often been used as a first
indicator how strongly alloy or even dot shape anisotropies affect
the FSS.~\cite{ScWi2007} To study in more detail the impact of the
alloy microstructure on the electronic structure,
Fig.~\ref{fig:randomelch6} depicts the energy eigenvalues of the
first three bound electron states ($e_0$, $e_1$, $e_2$) as a
function of the alloy configuration number. As shown in
Fig.~\ref{fig:randomelch6}, the above found large $p$-state
splitting in configuration 3 (Config. 3) is not an outlier; the
remaining four configurations produce very similar values for the
$p$-state splitting $\Delta E^\text{p,RA}_{e}$; $\Delta
E^\text{p,RA}_{e}$ varies between $\approx6-34$ meV depending on the
alloy microstructure.

\begin{figure}[]
    \centering
    \includegraphics[width=1\columnwidth]{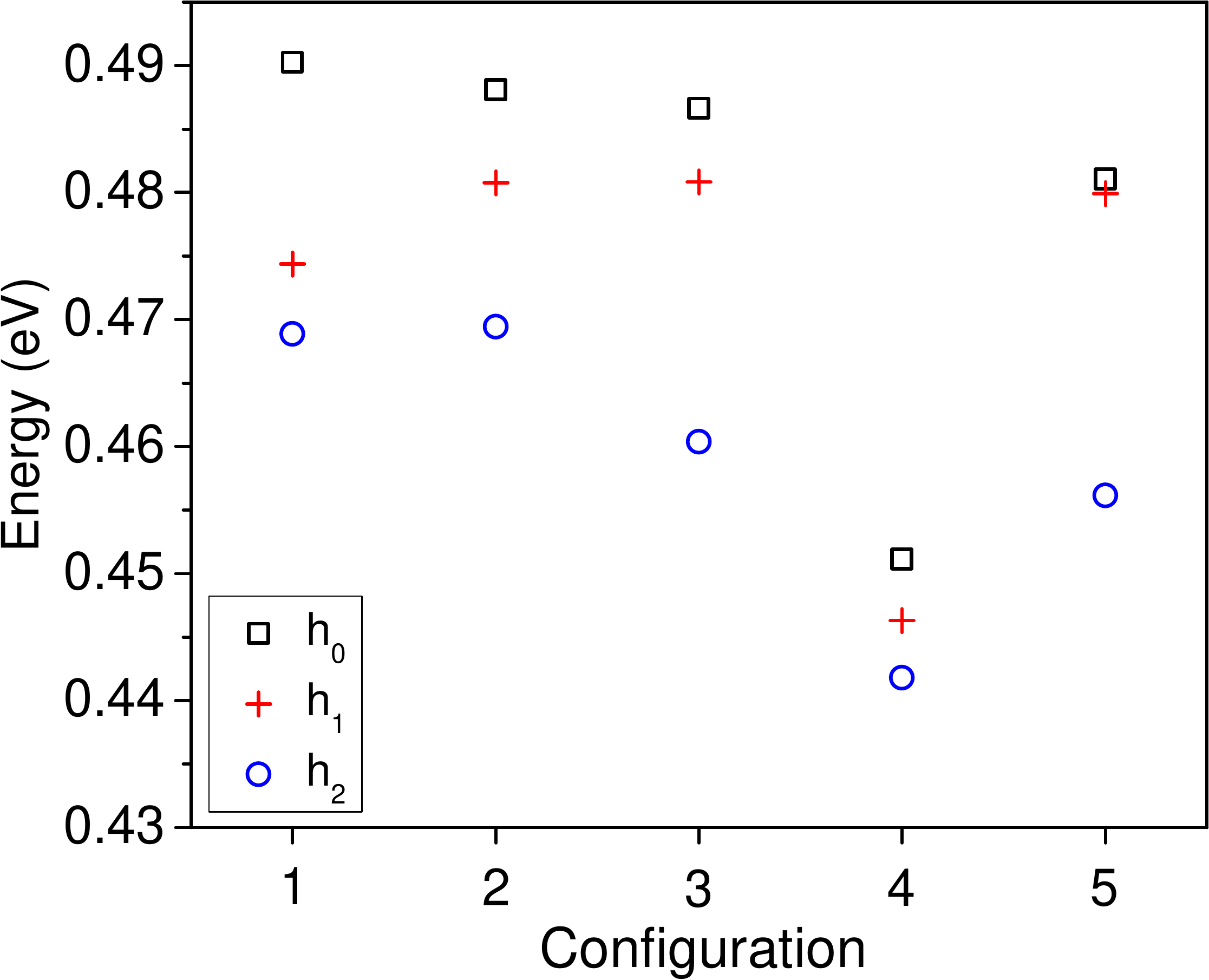}
    \caption{Hole ground, $h_0$, first $h_1$ and second $h_2$ excited
        state energies as a function of the microscopic alloy configurations
        number.}
    \label{fig:randomhoch6}
\end{figure}

Turning to the hole state energies $E^{i}_h$, data in
Fig.~\ref{fig:SP_states_holes} for alloy configuration 3 (Config. 3)
reveals that the first two hole states $h_0$ and $h_1$, which in the
VCA case are classified as $p$-like states, are split by $\Delta
E^\text{p,RA}_{h}\approx6$ meV; in the VCA case we find $\Delta
E^\text{p,VCA}_{h}=2.6$ meV. However, and as already discussed
above, while in the case of the electrons one may still want to
classify states according to their nodal structure as $s$- or
$p$-like, for the holes this classification is not possible.
Therefore, it is slightly more complicated to compare the splittings
between states directly. Nevertheless, in the following we use the
notation $\Delta E^\text{p,RA}_{h}$ to display the difference
between the first two hole states $h_0$ and $h_1$.
Figure~\ref{fig:randomhoch6} shows that $\Delta E^\text{p,RA}_{h}$
significantly depends on the alloy microstructure; we find that
$\Delta E^\text{p,RA}_{h}$ varies between $\approx1-15$ meV.

Overall, our calculations reveal that alloy fluctuations
significantly affect the electronic structure of $c$-plane InGaN/GaN
QDs. Our data also indicate that the symmetry of the system and as a
result, the symmetry of the wave functions is strongly affected.
This leads therefore to the question of how the excitonic fine
structure is impacted by random alloy fluctuations. Understanding
the excitonic fine structure will  allow us to understand the impact
of alloy fluctuations on entangled photon emission from $c$-plane
InGaN/GaN QDs. However, before turning to this question, we study
the excitonic fine structure and its consequences for entangled
photon emission in VCA.

\subsection{Excitonic fine structure splitting, excitonic binding energies and light polarization characteristics}
\label{sec:X_FSS}

Having discussed the electronic structure of $c$-plane InGaN/GaN QDs
in detail above, we focus our attention now on the optical
properties. Here, we look at the exciton fine structure, the exciton
binding energies, and the expected light polarization
characteristics of photons emitted from a $c$-plane InGaN/GaN QD.
All these quantities are important to understand when evaluating the
potential of $c$-plane InGaN/GaN QDs for entangled photon emission
at elevated temperatures. We start our analysis with the VCA
description. This will serve as a reference frame for the
calculations including random alloy fluctuations, discussed in
Sec.~\ref{sec:X_FSS_Atom}.

\subsubsection{Virtual crystal approximation results}
\label{sec:X_FSS_VCA}

In this section we discuss the optical properties of a $c$-plane
InGaN/GaN QD described within VCA. But, before looking at the
results from such a calculation, we apply group theory to gain
insight into the exciton fine structure and polarization
characteristics of the emitted photons. Overall, the group
theoretical analysis of the VCA data presents also a rigorous
benchmark of the theoretical framework.

Even though SOC effects are taken into account in our study, it is
informative to start with a discussion where this relatively small
contribution is neglected. In the case of a VCA and for the chosen
dot geometry the symmetry of the system is $C_{3v}$, as already
discussed in detail above. When neglecting SOC effects, the orbital
and spin part of a wave function can be decoupled and treated
independently. For the considered ideal $c$-plane WZ QD with
$C_{3v}$ symmetry, the electron ground state $e_0$ transforms like
an $s$-like state (see above) and corresponds to the IRR ${A_1}$ of
the $C_{3v}$ single group.~\cite{DuKa2011,BaSc2007} The $p$-like
hole ground state $h_0$ can be classified by the IRR
$E$.~\cite{DuKa2011,BaSc2007} It is important to note that $E$ is a
two-dimensional (2-D) IRR. Thus in the absence of SOC, the hole
ground state is two-fold degenerate (four-fold when including spin),
as for instance also found in the calculations of
Ref.~\cite{ScSc2008}.

Focussing on the orbital part of the states involved in
$e{_0}$-$h{_0}$ transitions, the corresponding exciton states
transform as $A_1\otimes{E}$ = $E$. This means, the exciton ground
state is two-fold degenerate since $E$ is a 2-D IRR. Taking now
electron and hole spins into account, the resulting states are
either singlet or triplet states.~\cite{Patra2020} Combining orbital
and spin parts of the excitonic wave function, and neglecting
initially also electron-hole exchange terms, the exciton ground
state is eight-fold degenerate. The electron-hole exchange
interaction lifts this degeneracy and one is left with a two-fold
degenerate state stemming from $E$ symmetry of the orbital part and
the 1-D $D_{0}$ IRR of the spin part and a six-fold degenerate state
stemming from $E$ symmetry of the orbital part and the 3-D $D_{1}$
IRR of the spin part.

Taking SOC effects into account, as done in our VCA calculations,
the single particle states transform according to the IRRs of the
$C_{3v}$ \emph{double} group. As discussed above, the hole ground
state $h_0$ has $p_x$- and $p_y$-like character and transforms
according to the 2-D $E_{3/2}$ representation. The electron ground
state $e_0$ has $s$-like character and transforms as the
two-dimensional IRR $E_{1/2}$. The exciton ground states transform
according to $E_{1/2}\otimes E_{3/2}$= $E\oplus E$. Since $E$ is a
2-D IRR of the $C_{3v}$ group, two two-fold degenerate excitonic
states are expected. Additionally, following Ref.~\cite{DuKa2011,
AlHe1994}, the two doublets are formed by bright states and the
excitonic emission spectrum is polarized in the $(x,y)$ plane, thus
perpendicular to $c$-axis ($z$-axis). In the presence of
electron-hole exchange interaction, the degeneracy of the two bright
$E$-symmetric states is lifted. We denote the splitting between
these two pairs of $E$-symmetric bright states as the
(bright-bright) FSS. We stress that the splitting between the bright
states within a $E$-symmetric doublet is zero. It is important to
note that this exciton fine structure is very different from a
$C_{2v}$-symmetric dot, e.g. an InGaAs/GaAs QD grown along the
[001]-direction of the ZB lattice. In a $C_{2v}$-symmetric system
the four lowest excitonic states consist of a dark state, a state
where the emitted photon is polarized along the QD growth direction
(e.g. [001]-direction) and two bright states, which emit photons
polarized within the growth plane ($x-y$-plane). The FSS of
[001]-oriented InGaAs/GaAs dots usually quoted in the literature, is
the splitting between the bright excitonic ground states. The
magnitude of this (bright-bright) splitting, along with the
radiative recombination lifetime determines the ability of a given
dot to emit entangled photon pairs, see
Eqs.~(\ref{eq:entangledstate1}) and~(\ref{eq:entangledstate2}). The
comparison between the excitonic fine structures in $C_{3v}$ and
$C_{2v}$ symmetric dots highlights now again very clearly the
benefit of targeting $C_{3v}$ symmetric dot systems: in this case
one is left with two doublets of bright exciton states which all
emit circularly polarized light. Therefore, $c$-plane III-N systems
should \emph{in principle} be ideal candidates for polarization
entangled photon pairs.

\begin{figure}[t!]
\centering
\includegraphics[width=1\columnwidth]{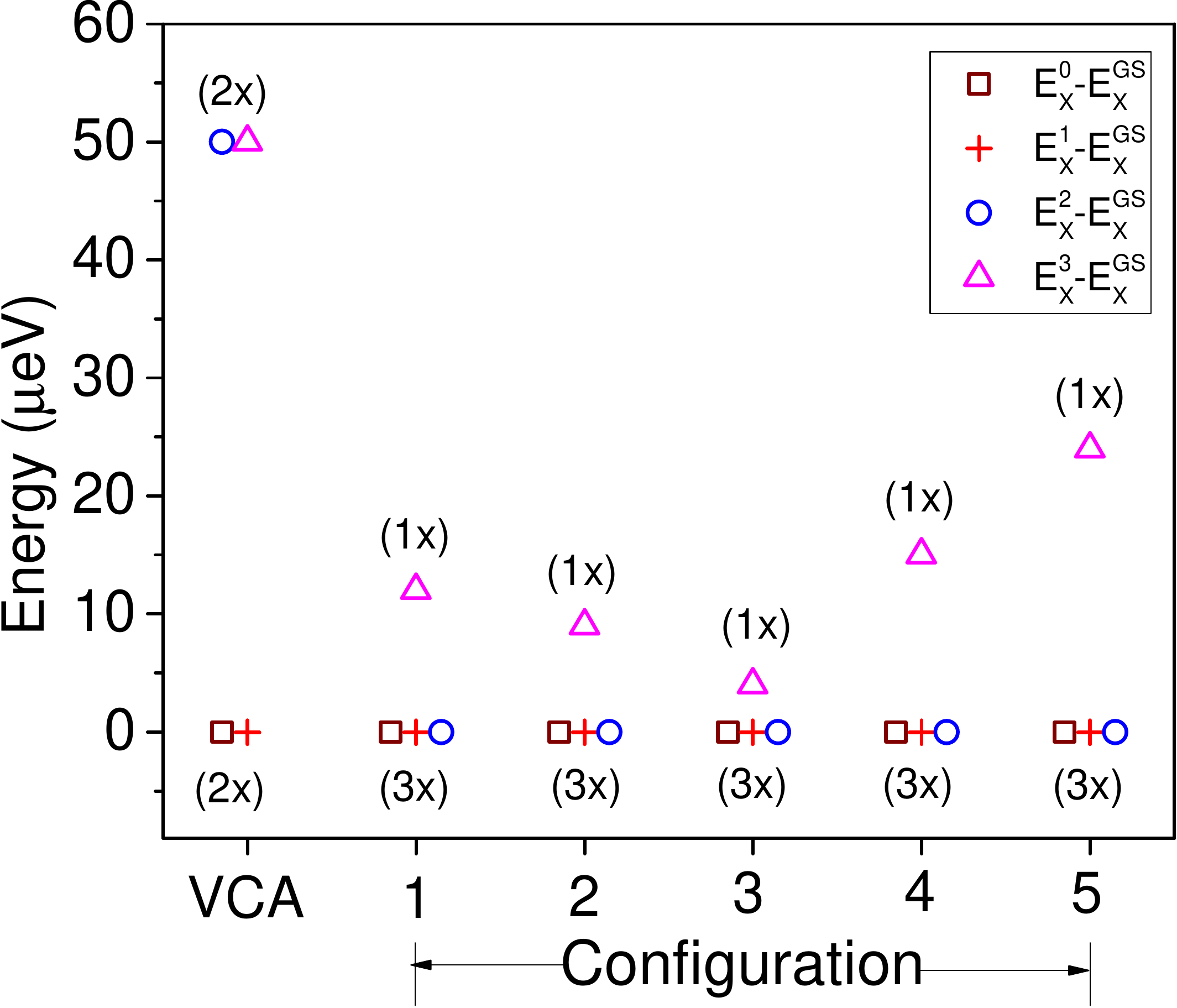}
\caption{Fine structure splitting of the four energetically lowest
exciton states in a $c$-plane In$_{0.2}$Ga$_{0.8}$N/GaN QD. The data
are shown for a virtual crystal approximation (VCA) of the dot and
when considering five different (random) alloy configurations. The
data is always displayed with respect to the exciton ground state
energy ($E^\text{GS}_X=E^0_X$).} \label{fig:FSSch6}
\end{figure}

Equipped with all this knowledge, we now present the outcome of our
VCA calculations for $c$-plane InGaN/GaN QDs. To compare the
calculated VCA FSS values with results from the fully atomistic
treatment (including alloy fluctuations), we proceed as follows: the
energies of the four energetically lowest lying exciton states,
$E^{i}_{X}$, are always plotted with respect to the ground state
energy, $E^\text{GS}_{X}$ ($E^{0}_{X}$) so that we are left with
$\Delta E^{i}_{X}=E_{X}^i-E^\text{GS}_{X}$, where $i\in\{0,1,2,3\}$.
The degeneracy of each level is indicated by $N$x, were $N$ gives
the degeneracy (e.g. 2x=two-fold degenerate exciton state). The
ground and the first three excited states are denoted by square,
cross, circle and triangle symbols, respectively.
Figure~\ref{fig:FSSch6} depicts $\Delta E^{i}_{X}$ for the VCA-case
and the random configurations, which will be discussed later.
Looking at the VCA data, our calculations show the group
theoretically predicted behavior of two sets of two-fold degenerate
exciton states. These levels are split by a (bright) FSS of
$\Delta_\text{FSS}\approx50$ $\si{\micro}$eV. We note that when
neglecting the electron-hole exchange interaction in the
calculations (thus only direct Coulomb terms are included) a
four-fold degenerate exciton ground state is found (not shown),
inline with our group theoretical analysis. Thus, the electron-hole
exchange interaction is responsible for the observed FSS. Moreover,
it is important to note that even though the SOC effect is weak in
InN and GaN systems, at least when compared to InAs or GaAs
systems,~\cite{VuMe2001,VuMe2003} without the SOC excitonic fine
structure looks completely different as discussed above in detail.

\begin{figure}
\centering
\includegraphics[width=0.65\columnwidth]{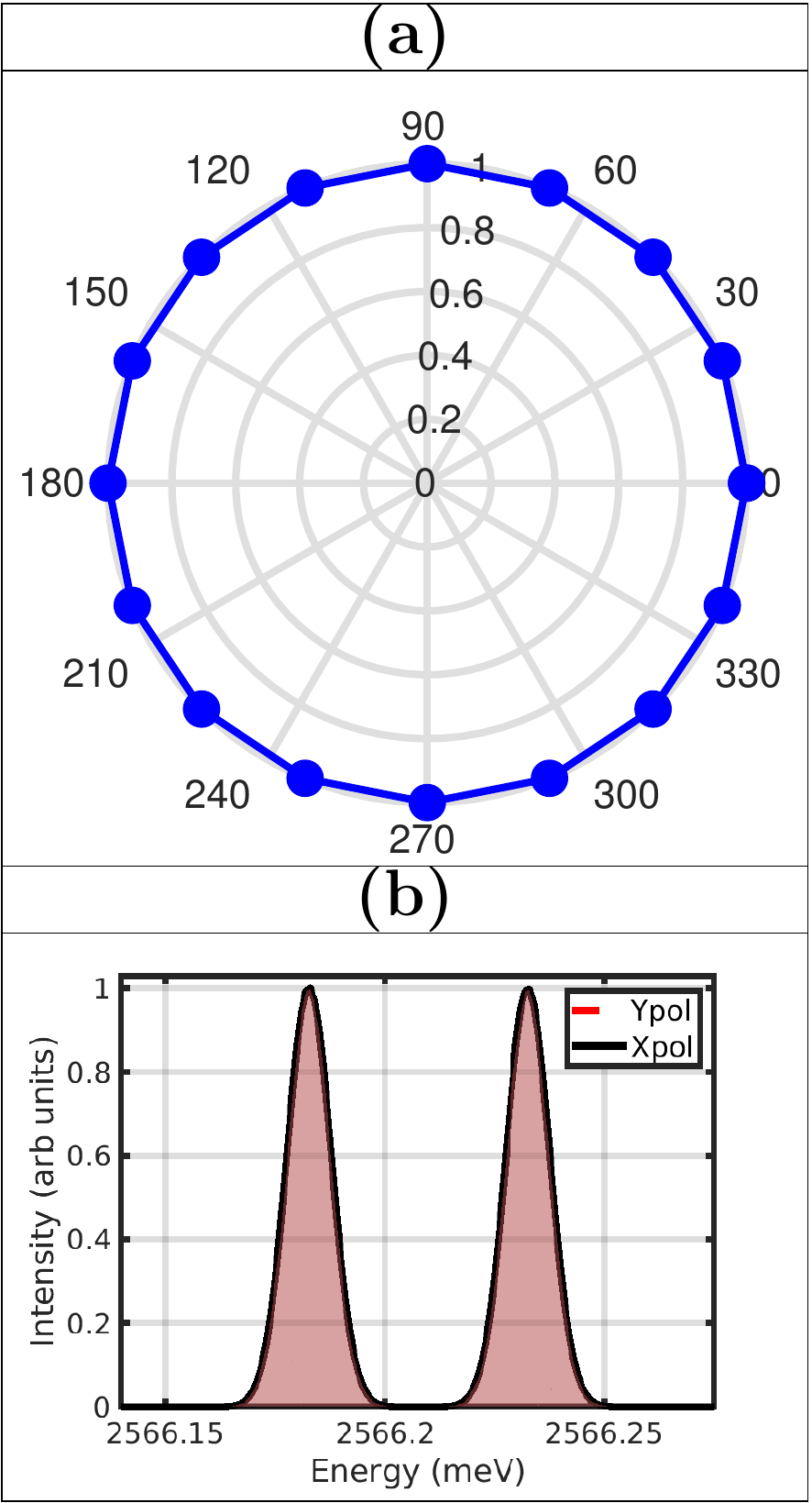}
\caption{(a) Polar plot of the optical intensity in the $c$-plane
($x-y$-plane) for the excitonic ground state emission of a $c$-plane
In$_{0.2}$Ga$_{0.8}$N/GaN QD when treated in a virtual crystal
approximation. (b) Excitonic absorption spectrum for two orthogonal
light polarization vectors. The light polarization vector along the
$x$-axis ($y$-axis) corresponds to $0^\circ$ ($90^\circ$) in the
polar plot of (a). Note that the $x$- and $y$-polarized absorption
are overlapping and thus can not be distinguished.}
\label{fig:VCAspectra}
\end{figure}

Having discussed excitonic fine structure we turn in a second step
to the light polarization characteristics of photons emitted/absoped
by the considered dots. To do so, we study the excitonic absorption
and emission spectra of the $c$-plane InGaN QD and compare the
calculated spectra with the group theoretical predictions.
Figure~\ref{fig:VCAspectra} (a) shows a polar plot of the optical
intensity of the exciton ground state emission. Here we have varied
the light polarization vector within the $x-y$-plane and the figure
reveals circularly polarized light, consistent with the group
theoretical analysis above; this behavior is ideally suited for
polarization entangled photon emission, see
Eq.~(\ref{eq:entangledstate1}). To shed further light onto the
optical properties of such a $C_{3v}$-symmetric dot,
Fig.~\ref{fig:VCAspectra} (b) displays the excitonic absorption
spectrum for two orthogonal light polarization vectors. Using the
polar plot from Fig.~\ref{fig:VCAspectra} (a), the $x$-polarized
light corresponds to an angle of $0^\circ$, while $y$-polarized
light corresponds to an angle of $90^\circ$.
Figure~\ref{fig:VCAspectra} (b) reveals, independent of the light
polarization, two distinct absorption lines are separated by 50
$\mu$eV. Note that the $x$- and $y$-polarized absorption are
overlapping and thus can not be distinguished. The splitting between
the two absorption peaks is inline with the excitonic fine structure
discussed above and shown in Fig.~\ref{fig:FSSch6}. Overall, the
results are in accordance with the group theoretical prediction that
(i) the doublets present in the excitonic fine structure are bright
and (ii) the emitted light is polarized in the $(x-y)$ plane.
Therefore, photons emitted via these transitions can be classified
as ($\sigma^{-}$) or right ($\sigma^{+}$) circularly polarized.

We stress again, that the FSS of $\Delta_\text{FSS}=50$
$\si{\micro}$eV between bright excitonic doublets is of secondary
importance for entangled photon emission. This stems from the fact
that the FSS of the two bright states per doublet is zero and the
light emitted is circularly polarized. As a result, in the XX-X exciton
cascade involving these bright doublet states, one is left with the
situation schematically depicted in
Fig.~\ref{fig:finestructuresplitting} (a) where due to the
degeneracies of the bright exciton ground states, both left and
right arms of the cascade emit circularly polarized light.
Therefore, a polarization entangled state given by
Eq.~(\ref{eq:entangledstate1}) can be obtained. We highlight again
that all this originates from the underlying $C_{3v}$-symmetry of
the system, so that this analysis is also applicable to other system
($c$-plane GaN/AlN; [111]-oriented InAs/GaAs QDs) which
exhibits and preserves a $C_{3v}$ symmetry.

\begin{figure}[t!]
\centering
\includegraphics[width=1\columnwidth]{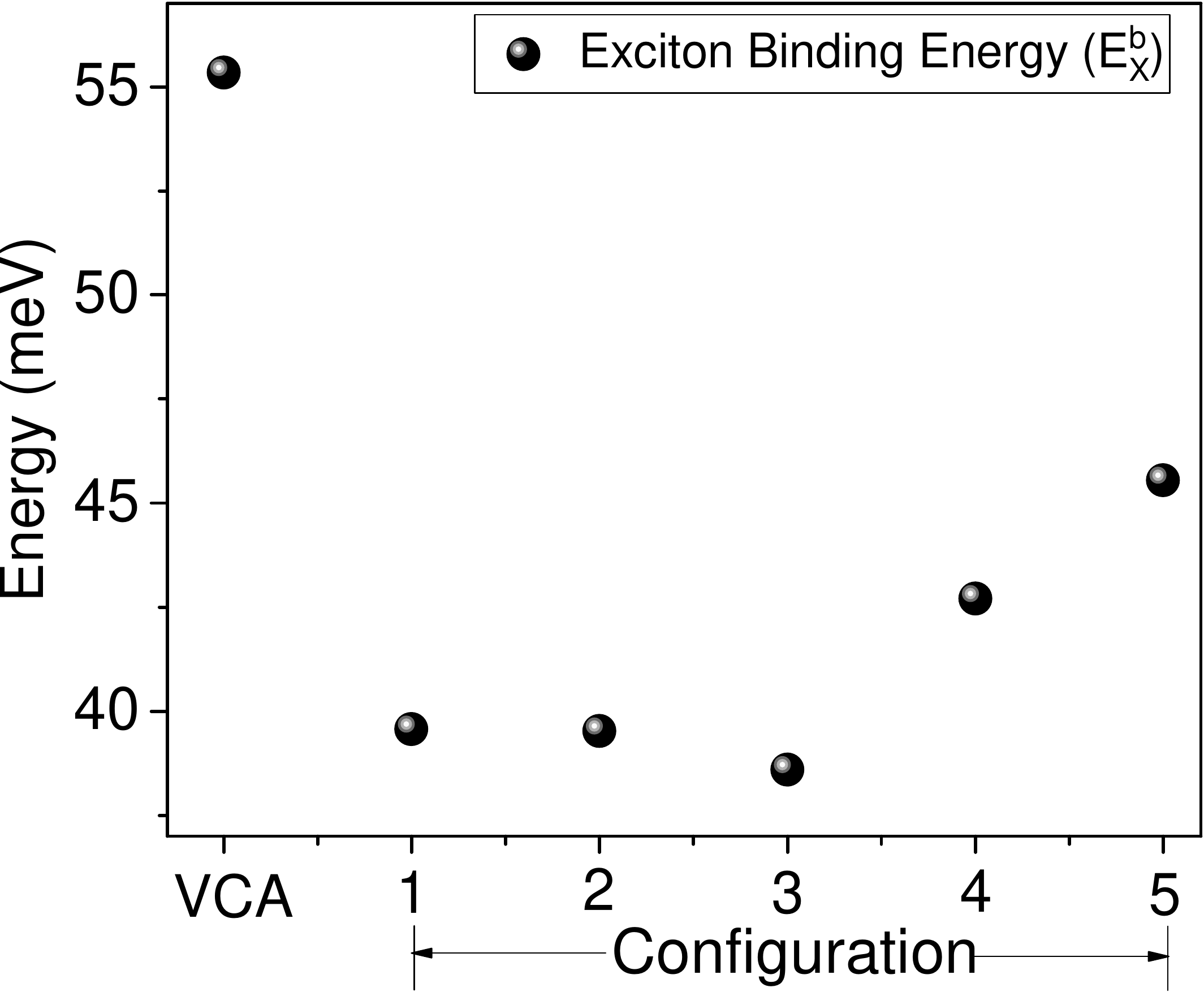}
\caption{Excitonic binding energy $E_{X}^{b}$ of $c$-plane
In$_{0.2}$Ga$_{0.8}$N/GaN QDs. The data are shown for the virtual
crystal approximation (VCA) case and for the five different random
alloy configurations.} \label{fig:bindingenergy}
\end{figure}

While all these aspects already highlight the benefit of a $C_{3v}$
symmetric dot for polarization entangled photon emission, WZ III-N
$c$-plane dots should ideally offer, in part thanks to the large
band offsets, excitonic binding energies higher than the thermal
energy at room temperature. Thus, in that case polarization
entangled photon emission at much higher temperatures as in the
arsenide systems can be achieved. Thus, we have also analyzed the
excitonic binding energy $E_{X}^{b}$ of the considered $c$-plane
In$_{0.2}$Ga$_{0.8}$N/GaN dot. The exciton binding energy is
calculated via $E_{X}^{b}$ = ($E^{0}_{e}-E^{0}_{h}-E_{X}$), where
$E^{0}_{e}$ and $E^{0}_{h}$ are the electron and hole
single-particle ground state energies, respectively, while $E_{X}$
is the exciton ground state energy. Figure~\ref{fig:bindingenergy}
depicts $E_{X}^{b}$ for the VCA case and the five different random
alloy configurations. For the VCA case we obtain a very large
exciton binding energy of $\approx$ 55 meV, which clearly exceeds
the thermal energy at room temperature of $\approx$ 25 meV. Overall,
and taking the large excitonic binding energy, excitonic fine
structure and the light polarization characteristics of the emitted
photons into account, our study shows that $c$-plane InGaN QDs are
in principle ideal candidates for entangled photon emission at
elevated temperatures.

Having discussed the \emph{ideal} VCA case we turn now and focus our
attention on the impact of random alloy fluctuations on the optical
properties of $c$-plane InGaN/GaN QDs. As above, special attention
will be paid to impact of such alloy fluctuations on entangled
photon emission from such dots.

\subsubsection{Random alloy fluctuation results}
\label{sec:X_FSS_Atom}

We address in this section the consequences of random alloy
fluctuations in $c$-plane InGaN/GaN QDs on their ability to emit
polarization entangled photon pairs. Before presenting the results
of our CI calculations, similarly to the VCA case, we start with a
group theoretical analysis.

Building on our discussion of the single-particle states in
Sec.~\ref{sec:SPstates_energies_atomistic}, we have seen that the
$C_{3v}$-symmetry is lost in the presence of random alloy
fluctuations. As we have already mentioned in
Sec.~\ref{sec:SPstates_energies_atomistic}, we are left with a $C_1$
symmetry. Again neglecting initially the weak SOC, the electron
$e_0$ and hole $h_0$ ground states transform according to the IRR
$A$ of the $C_1$ single group. Thus the exciton state related to the
$e_0$-$h_0$ transition has $A\otimes A=A$ symmetry, which is 1-D
IRR. When including spin, but neglecting SOC and electron-hole
interaction in the description, the exciton ground state is
four-fold degenerate. The electron-hole exchange interaction splits
this four-fold degenerate state into singlet and triplet states.
Accounting for SOC, electron and hole ground states transform
according to the $A_{1/2}$ IRR of the $C_1$ double group. In this
case, the excitonic states transform according to $A_{1/2}\otimes
A_{1/2}=A$. Since $A$ is a 1-D IRR, one expects four non-degenerate
states for the four energetically lowest exciton states.

Figure~\ref{fig:FSSch6} shows the exciton fine structure for the
five different random alloy configurations. From this figure one can
infer that when taking random alloy fluctuations into account the
excitonic FSS is very different from the VCA case. While in VCA one
observes two doublets of degenerate states, in the random alloy case
we find a non-degenerate state and an (approximately) three-fold
degenerate exciton state; this finding is independent of the
configuration number, thus the alloy microstructure. Overall,
bearing in mind our group theoretical analysis: for a system with
$C_1$-symmetry, including SOC and electron-hole exchange
interaction, the energetically lowest lying excitonic states are
expected to be non-degenerate. Here, our calculated excitonic fine
structure is similar to the case expected in the absence of SOC
effects, where due to electron-hole exchange interaction a singlet
(non-degenerate) state and a triplet (three-fold degenerate) state
is expected. While one might be tempted to classify the calculated
excitonic states as being singlet and triplet states, we note that
the three states forming the ``triplet'' states are only
approximately degenerate, indicating that SOC starts to mix singlet
and triplet states as observed in other systems.~\cite{BeSh2001} The
small FSS between ``singlet'' and ``triplet'' states of 5-25 $\mu$eV
can facilitate such a mixing. The small FSS calculated in our
$c$-plane system is attributed to the presence of the strong
electrostatic built-in field ($\approx$ 0.09 MV/cm)~\cite{VeJe2014}
in $c$-plane InGaN/GaN QDs along the $c$-axis. This field spatially
separates electron and hole wave functions along the growth
direction of the dot and reduces therefore the electron-hole
exchange Coulomb matrix elements in comparison to system where such
a strong field is absent, e.g. [001]-oriented InGaAs/GaAs dots.
Therefore, the combination of SOC and small electron-hole exchange
interaction may lead to singlet-triplet state mixing. Before looking
at light polarization characteristics and how this quantity is
affected by random alloy fluctuations we first discuss the exciton
binding energies. As shown in Fig.~\ref{fig:bindingenergy}, and
similar to the VCA case, we observed large excitonic binding energy
values ranging between 38-46 meV. Therefore, excitons in these
systems are expected to be stable near room temperature irrespective
of the alloy microstructure, since $E^{b}_X$ still exceeds the
thermal energy at room temperature.

Having discussed the excitonic fine structure and the exciton
binding energies, we target now the light polarization
characteristic of the emitted photons. As highlighted already above,
not only the FSS is important but also circularly or orthogonally
polarized photons are required to achieve polarization entangled
photon pairs. In the following we use the random alloy configuration
5 (Config. 5) as an example, since it exhibits the largest FSS (cf.
Fig.~\ref{fig:FSSch6}). We note here that all results discussed
below for this selected configuration reflect the behavior found in
the four remaining alloy configurations.

\begin{figure}[t!]
\centering
\includegraphics[width=0.65\columnwidth]{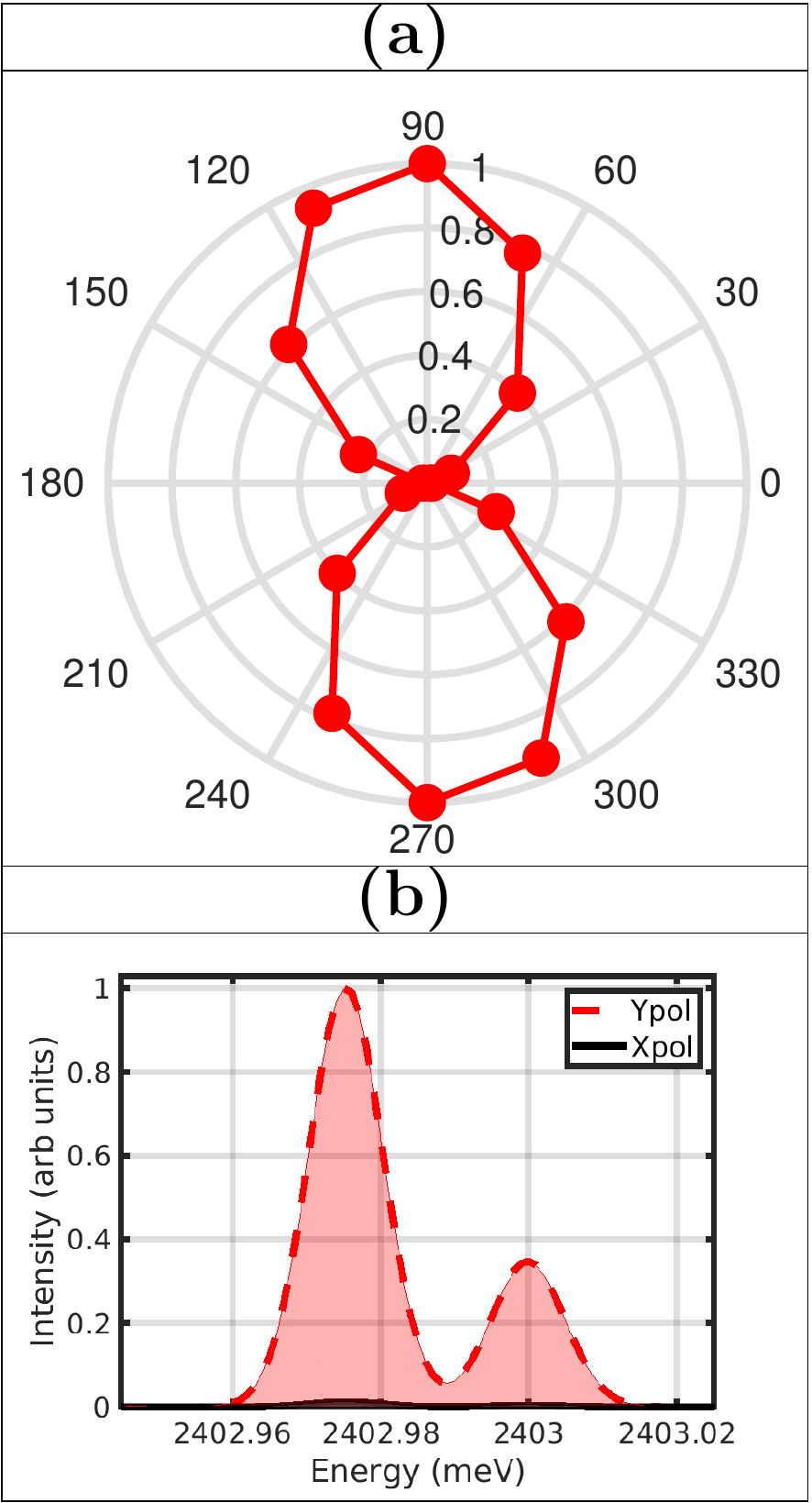}
\caption{(a) Polar plot of the optical intensity in the $c$-plane
($x-y$-plane) of the excitonic ground state emission of a $c$-plane
In$_{0.2}$Ga$_{0.8}$N/GaN QD when accounting for random alloy
fluctuations. (b) Excitonic absorption spectrum for two orthogonal
light polarization vectors. The light polarization vector along the
$x$-axis ($y$-axis) corresponds to $0^\circ$ ($90^\circ$) in the
polar plot displayed in (a).} \label{fig:Randomspectra_2}
\end{figure}

\begin{figure}[t]
\centering
\includegraphics[width=0.4\columnwidth]{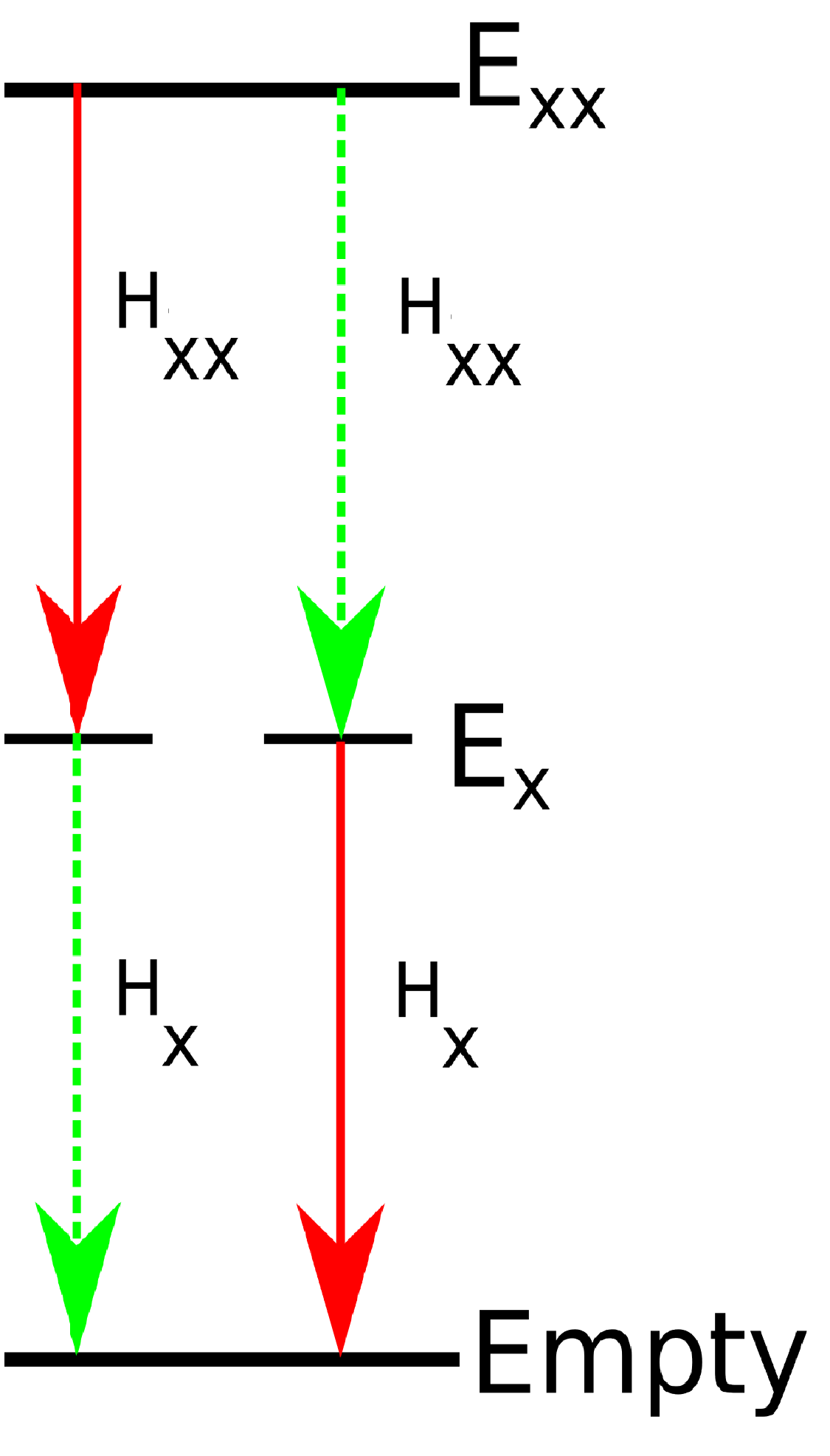}
\caption{Schematic illustration of the light polarization properties
of the emitted photons from biexciton-exciton cascade in a $c$-plane
InGaN/GaN QD when including random alloy fluctuations. Only two
states from the three-fold degenerate states are selected. The
emitted photons are horizontally polarized to a given reference
axis.} \label{fig:finestructuresplitting2}
\end{figure}

Figure~\ref{fig:Randomspectra_2} (a) depicts a polar plot of the
optical intensity of the exciton ground state emission for Config.
5. The angle dependence of optical intensity is again analyzed
within the $c$-plane. Comparing this result with the data from the
VCA calculation, Fig.~\ref{fig:VCAspectra}, the difference is
striking. While in the VCA case one is left with circularly
polarized light, in the random alloy cases the light is linearly
polarized. We note that we find strongly linearly polarized light
for all other random alloy configurations. As discussed above,
linearly polarized light emission and a non vanishing FSS is not
necessarily a problem for entangled photon emission as long as (i)
the FSS is not larger than the linewidth, (ii) the radiative
lifetime is small and (iii) orthogonally polarized photon emission
is possible from the XX-X cascade. To shed light onto aspect (iii),
Fig.~\ref{fig:Randomspectra_2} (b) displays the excitonic absorption
spectra for Config. 5. The absorption spectra reveals two peaks,
which are basically separated by the FSS found for Config. 5 (cf.
Fig.~\ref{fig:FSSch6}), thus reflecting the excitonic fine structure
discussed above. We note also that if the calculated states were
triplet and singlet states, transitions involving the triplet states
should have been dark. However, as Fig.~\ref{fig:Randomspectra_2}
(b) reveals this is not the case. Therefore, we conclude that the
small electron and hole exchange interaction in conjunction with the
presence of the SOC leads to a singlet and triplet state mixing. As
a consequence the energetically lowest four excitonic states are all
bright. At first glance, the fact that all four excitonic states are
bright and that the FSS is relatively small may still be compatible
with generating entangled photon pairs. The light polarization
characteristics, however, are a major drawback for
\emph{polarization} entangled photon emission in the case of random
alloy fluctuations. Looking back at aspect (iii), emission of
orthogonally polarized photons is required to generate such a photon
pair. But as Fig.~\ref{fig:Randomspectra_2} (b) reveals, photons
generated from transitions involving the different bright excitonic
states have the same polarization characteristics; the situation is
schematically illustrated in in
Fig.~\ref{fig:finestructuresplitting2}. Using the notation from
Fig.~\ref{fig:finestructuresplitting2}, only states of the form
\begin{equation}
|\psi\rangle=\frac{1}{\sqrt{2}}\left(\left|H_{XX}^{}
H_{X}^{}\right\rangle+\left|H_{XX}^{}
H_{X}^{}\right\rangle\right)\,\, \,
\end{equation}
can be realized. Such a state is \emph{not} a polarization entangled
state. Therefore, even though we have a three-fold degenerate bright
excitonic state (vanishing FSS), the polarization characteristics of
the emitted photons will prohibit polarization entanglement, in the
case that the dot exhibits random alloy fluctuations.

The (in-plane) light polarization characteristics of excitonic
transitions in $c$-plane InGaN/GaN QDs has also been studied
experimentally.~\cite{AmMo2012,JeMa2015,ChKi2018} Amloy~\emph{et
al.}~\cite{AmMo2012} for instance reports strongly linearly
polarized excitonic emission spectra from single $c$-plane InGaN/GaN
QDs with 20\% In content. In Refs.~\cite{JeMa2015,ChKi2018} also
strongly linearly polarized excitonic transitions from InGaN/GaN QDs
have been reported.  Overall, these findings are inline with the
here reported theoretical results.

\section{Conclusion}
\label{sec:Conclusion}

In this study we have analyzed the potential of $c$-plane InGaN/GaN
dots for polarization entangled photon emission at elevated
temperatures. Special attention was paid to consequences of random
alloy fluctuations for this question. The analysis was carried out
in a fully atomistic many body framework accompanied by group
theoretical considerations.

Using virtual crystal approximation calculations as a starting
point, thus neglecting alloy fluctuations entirely, we find an
excitonic fine structure that is very different from the fine
structure in more conventional III-V-based quantum dot systems but
ideally suited for entangled photon emission. Also the polarization
characteristics of the emitted photons meet the requirements for
polarization entanglement. Additionally, our analysis reveals a very
large excitonic binding energy ($\approx$50 meV). This large binding
energy exceeds the thermal energy at room temperature and should
thus support stable excitons near room temperature. Overall, our
studies on these \emph{idealized} structures indicate that $c$-plane
InGaN/GaN quantum dots are highly attractive for high temperature
entangled photon emission.

While the benefit of large excitonic binding energies is still
preserved when accounting for random alloy fluctuations on a
microscopic level, the excitonic fine structure is significantly
modified when compared to the virtual crystal approximation. Our
calculations show that a major drawback of this modified fine
structure stems from the fact that while energetically lowest
excitonic states are all bright and degenerate, the photons emitted
from a biexciton-exciton cascade involving these states exhibit
basically the same polarization characteristics. This aspect thus
prevents polarization entangled photon emission via the
biexciton-exciton cascade in $c$-plane InGaN/GaN quantum dots.

We note that our underlying assumption is a random alloy description
of the quantum dot microstructure. If the alloy composition can be
manipulated, for instance by introducing In atom clustering as for
instance observed in non-polar InGaN systems,~\cite{TaZh2019} there may be
ways to tailor the alloy microstructure so that the system exhibits
again a $C_{3v}$ symmetry. In doing so the full potential of $c$-plane
InGaN/GaN dots may be exploited for polarization entangled photon
pairs.

Going beyond polarization entanglement, photon entanglement may be
realized by using other approaches such as the arrival time of a
single photon~\cite{JaPr2014}. Time-binned entangled photon pairs
have been realized recently by employing coherent superposition of
``early'' and ``late'' photons in a biexciton-exciton
cascade;~\cite{JaPr2014} this scheme does not rely on the
polarization characteristics of the emitted photons. Finally, twin
photon emission, where actually photons of the same polarization are
required, provide a further avenue for non-classical light
emission.~\cite{PaSc2020} All these schemes, which have mainly been
addressed for arsenide-based system, should benefit from the here
observed large excitonic binding energies and in general the large
band offset. Therefore $c$-plane III-N based QD system should
provide interesting properties and routes to achieve non-classical
light emission at elevated temperatures over a wide emission
wavelength range.

\ack

The authors thank the Science Foundation Ireland (project nos.
13/SIRG/2210, 17/CDA/4789, and 12/RC/2276 P2) for the financial
support of this project and Emanuele Pelucchi for fruitful
discussions. Furthermore,computing resources provided by Science
Foundation Ireland (SFI) to the Tyndall National Institute and
Higher Education Authority funded Irish Centre for High End
Computing (ICHEC) are acknowledged.

\appendix

\section*{Appendix} \label{sec:Appendix}

In this section we provide insight into theory underlying the
results presented in the main text of the manuscript. We will start
with the electronic structure theory applied,
Sec.~\ref{sec:Electronic_Struc_Theory}. The details of the
configuration interaction calculations are given in
Sec.~\ref{sec:Many_Body_calculations}.

\section{Electronic structure theory}
\label{sec:Electronic_Struc_Theory}

To gain insight into the electronic structure of $c$-plane InGaN/GaN
QDs we have employed a nearest neighbor $sp^3$
model~\cite{ScSc2006}. In general, this model has already been
discussed in detail in our previous work, targeting bulk and quantum
well structures. For these systems it shows very good agreement with
available experimental data for a variety of different
aspects.~\cite{CaSc2013,ScCa2013Apex,CoSc2015,TaCa2016,TaDa2020}
Overall, this model presents now an ideal starting point for
calculating the electronic structure of InGaN/GaN QDs on a
microscopic level. We discuss in the following how the model is
adjusted to describe the electronic structure of QDs and, in a
second step, their excitonic fine structure. While the above
discussed TB model is intrinsically an atomistic approach, we detail
also how it is modified to obtain a VCA model that still keeps the
underlying atomistic symmetry. We first start with the fully
atomistic approach, Sec.~\ref{sec:Theory_FA} before turning to the
VCA description, Sec.~\ref{sec:Theory_VCA}.

\subsection{Fully atomistic model} \label{sec:Theory_FA}

Our nearest neighbor TB model has been parametrized using
(Heyd-Scuseria-Ernzerhof) hybrid functional density functional band
structure data for GaN and InN.~\cite{CaSc2013} This model gives a
very good description of the unstrained bulk band structures near
the $\Gamma$-point but also at the edges of the first Brillouin zone
of lowest conduction and highest valence bands of both InN and GaN.
We have extended this parameterization here to include SOC
effects.~\cite{PaSc2020} Even though the SOC is small in InN
($\Delta^\text{InN}_\text{SOC}=$ 5 meV) and GaN
($\Delta^\text{GaN}_\text{SOC}=$ 17 meV), at least when compared to
InAs ($\Delta^\text{InAs}_\text{SOC}=$ 390 meV) and GaAs
($\Delta^\text{GaAs}_\text{SOC}=$ 341 meV), these effects are
important for accurate description of the excitonic fine structure.
SOC effects are included in the model by the widely used approach of
Chadi~\cite{Chadi1977}. Given the small values of the SOC, the
refitted TB parameters when including SOC effects, summarized in
Table~A1, differ only slightly from those in the absence of SOC.
Also, given that we are mainly interested in the SOC splitting of
the bands near the $\Gamma$-point, we do not distinguish between the
anion and cation based SOC TB parameters $\lambda_i$; distinguishing
between $\lambda_c$ and $\lambda_a$ would only result in the
situation that splittings of states lying energetically far away
from the band edges would be modified differently. However, these
energetically remote bands are of secondary importance for the
direct band gap materials InN and GaN, where bands close to the band
edge are mainly relevant for optical properties of InGaN/GaN QDs. To
treat alloy effects in this fully atomistic framework, we use for
the anion sites composition weighted averages of the TB parameters
describing the binary materials; such an approach has also been
widely applied by other groups.~\cite{LiPo92,OReLi2002,BoKh2007}
This averaging procedure is not required for cations, given that
their nearest neighbors are always nitrogen atoms. Furthermore, to
align the energy scales between InGaN and GaN, we then use natural
band offsets from HSE-DFT calculations.~\cite{MoMi2011}

\begin{table}
    \small
    \label{TBparameters}
    \centering \caption{Tight-binding parameters (in eV) for the nearest neighbor $sp^3$ model of wurtzite InN and GaN with ($\Delta_{\text{so}}\neq0$) and without ($\Delta_{\text{so}}=0$) spin-orbit coupling effects. The notation of Ref.~\cite{KoSa1983} is used.}
    \begin{tabular}{|l|c|c||c|c|}
        \hline\hline  & \multicolumn{2}{|c||}{$\Delta_{\text{so}}=0$} & \multicolumn{2}{c|}{$\Delta_{\text{so}}\neq0$} \\
        \hline
        & InN [eV] & GaN [eV] & InN [eV] & GaN [eV]\\
        \hline
        E(s,a) & -11.9173 & -10.6158 & -11.9173 & -10.6158\\
        E(p,a) & 0.4886 & 0.8183 & 0.4867 & 0.8127\\
        E($\text{p}_z$,\text{a}) & 0.4558 & 0.7926 & 0.4572 & 0.7849 \\
        E(s,c) & 0.4837 & 0.9122 & 0.4837 & 0.9122 \\
        E(p,c) & 6.5322 & 6.6788 & 6.5322 & 6.6788 \\
        V(s,s) & -1.6124 & -5.9749 & -1.6124 & -5.9749 \\
        V(x,x) & 1.7863 & 2.3381 & 1.7863 & 2.3381 \\
        V(x,y) & 4.8338 & 5.4697 & 4.8338 & 5.4697 \\
        V(sa,pc) & 1.8919 & 4.0909 & 1.8919 & 4.0909\\
        V(pa,sc) & 6.1355 & 8.6655 & 6.1355 & 8.6655 \\
        $\lambda_a$ & 0 & 0 & 0.0017 & 0.0052 \\
        $\lambda_c$ & 0 & 0 & 0.0017 & 0.0052 \\
        \hline\hline
    \end{tabular}
\end{table}

To study heterostructures, also strain effects have to be included
in the model. To do so, deformation potentials from HSE-DFT
calculations~\cite{MoMi2011} have been used and included in the
framework via a site-diagonal correction as detailed in
Ref.~\cite{ScCa2015}. In doing so the deformation potentials are
input parameters and no further fitting is required. To obtain the
relaxed atomic positions, we apply here Martin's valence force field
(VFF) model, which includes electrostatic interactions in the system
explicitly and reproduces characteristics of a real WZ system, for
instance, $\frac{c}{a}$ ratios, accurately.~\cite{Ma1970,CaSc2013}
More details on the model can be found elsewhere.~\cite{ScCa2015,
CaSc2013}

Finally, in WZ (local) strain effects will also give rise to very
strong (local) piezoelectric polarization fields, which in $c$-plane
InGaN/GaN QDs result in strong spatial separation of electrons and
holes along the growth direction. In addition, and different to
[001]-oriented ZB InGaAs/GaAs, WZ III-N materials exhibit also a
spontaneous polarization that is even present in the absence of
strain effects. To achieve a microscopic description of these
polarization effects, we employ here our recently developed local
polarization theory. So far this model has been applied to \emph{QW}
systems and care must be taken when employing it in a \emph{QD}
situation. In our local polarization theory the $i$th component of
the polarization vector field $\mathbf{P}$ can be written
as:~\cite{CaSc2013}
\begin{eqnarray}
\label{eq:macro}
P_{i}&=&\underbrace{\sum_{j=1}^{6} e_{i j}^{(0)}
    \epsilon_{j}}_{\text {macroscopic
}}\\\nonumber
& &+\underbrace{P_{i}^{\mathrm{sp}}-\frac{e}{V_{0}}
    \frac{Z_{i}^{0}}{N_{\mathrm{coor}}^{0}}\left(\mu_{i}-\sum_{j=1}^{3}\left(\delta_{i
        j}+\epsilon_{i j}\right) \mu_{j, 0}\right)}_{\text {local }}
\label{eq:local}
\end{eqnarray}
For the local part, Eq.~(\ref{eq:local}), that contains the
spontaneous polarization $P_{i}^{\mathrm{sp}}$, the bond asymmetry
parameter $\mu$, defined as a summation over the nearest neighbor
distances, the asymmetry parameter of the unstrained system $\mu_0$,
the strain tensor $\epsilon$, the Born effective charges $Z$, the
number of nearest neighbors ${N_{\mathrm{coor}}^{0}}$, and the
elementary charge $e$,  no changes are required to treat a QD
instead of a QW. The macroscopic part, Eq.~(\ref{eq:macro}), that
contains the clamped ion contribution, the situation is different.
In a QW system, the macroscopic part is zero in the barrier material
(e.g. GaN) but nonvanishing and constant in the well (e.g. InGaN).
This results from the fact that the barrier is unstrained and in the
well the macroscopic strain is constant. However, this does not hold
for a QD where the strain field is position dependent and also
non-vanishing in the barrier material close to the dot. To take this
into account, we apply a surface integral approach to calculate on
our atomistic (ideal WZ) grid the strain field from:~\cite{WiAn2005}
\begin{equation}
\epsilon_{i j}(\mathbf{r})=\delta_{i j} \epsilon_{0} \chi_{Q
    D}+\frac{\epsilon_{0} A}{4 \pi} \int_{Q D}
\frac{\left(x_{i}-x_{i}^{\prime}\right)}{\left|\mathbf{r}-\mathbf{r}^{\prime}\right|^{3}}
\hat{\mathbf{n}}_{j} \cdot d \mathbf{S}^{\prime}
\label{eq:SI_strain}
\end{equation}
where the primed quantities refer to points on the surface of the
dot, $(x_1,x_2,x_3)\equiv(x,y ,z)$, $\hat{\mathbf{n}}_j$ is the unit
vector in the $j$-direction, and $A=(1+\nu)/(1-\nu)$, with $\nu$
being the Poisson ratio. The characteristic function $\chi_{QD}$ is
equal to 1 inside the dot and zero outside. The isotropic misfit
strain is denoted by $\epsilon_{0}$ and assumed equal to
$\frac{1}{3}(2\epsilon_{0,a}+\epsilon_{0,c})$, where
$\epsilon_{0,a}$ is the misfit strain in the $c$-plane, and
$\epsilon_{0,c}$ is the misfit strain along the $c$-axis. This
analytic model allows us to calculate in a continuum-based frame on
our atomistic grid the clamped ion contribution required in the
local polarization theory, Eqs.~(\ref{eq:macro})
and~(\ref{eq:local}). Once the local polarization vector field is
known we use a point dipole method to calculate the corresponding
built-in potential.~\cite{CaSc2013} This (local) built-in potential
is included in the TB model as a diagonal correction, which is a
widely used ansatz to electrostatic built-in fields in a TB
model.~\cite{Schulz2007}

Having discussed the fully atomistic framework to study the
electronic structure of InGaN/GaN QDs, and to expose the impact of
alloy fluctuations on their electronic structure and finally optical
properties, we use a VCA model as a reference system. In the next
section we briefly describe how we construct such a model while
still keeping the underlying $C_{3v}$ symmetry of the combined
system of WZ crystal structure and QD geometry.

\subsection{Virtual crystal approximation} \label{sec:Theory_VCA}

To obtain a VCA-TB description of an alloyed In$_{x}$Ga$_{1-x}$N QD
system, the required TB parameters are the composition weighted
averages of the InN and GaN binary TB parameters. We note that the
chosen approach not necessarily agrees with the band gap bowing
parameter obtained from a fully atomistic description, however, the
transition energy for our study here is of secondary importance
since the VCA-TB model is designed to keep the perfect $C_{3v}$
symmetry of the system. Therefore, we use the VCA-TB parameters and
assume an ideal WZ lattice: in the barrier material we use the GaN
TB parameters while in the QD region virtual InGaN atoms described
by VCA-TB parameters. For a comparison with the fully atomistic
approach, an In content of 20\% has been assumed. To obtain the
strain field in and around the dot, we use the surface integral
approach discussed above, Eq.~(\ref{eq:SI_strain}). This gives an
easy and straightforward approach to obtain a VCA compatible strain
field that still keeps the $C_{3v}$-symmetry of the system. The
additional benefit of Eq.~(\ref{eq:SI_strain}) is that it assumes an
infinite matrix surround the dot. Therefore the rectangular boundary
conditions of the simulation cell are not a problem for this
approach.

Finally, to obtain the electrostatic built-in potential, arising
from spontaneous and piezoelectric polarization, we also use a
surface integral technique.~\cite{WiAn2005} This method is a simple
real space model that for certain dot geometries (e.g. cuboid) can
give analytic results for polarization potential in and around a
dot. This method uses surface integrals to evaluate these values and
assumes that the QD is buried in an infinite system; therefore,
boundary condition will not play a role. Since this method can be
used with any arbitrary underlying grid, this is attractive for use
in our TB model. The piezoelectric polarization potential is
obtained via:

\begin{eqnarray}
\varphi_{s t r}(\mathbf{r})&=&J \int_{Q D}
\frac{\left(x_{3}-x_{3}^{\prime}\right)^{2}}{\left|\mathbf{r}-\mathbf{r}^{\prime}\right|^{3}}
\hat{\mathbf{n}}_{3} \cdot d \mathbf{S}^{\prime}\\\nonumber
& & +K \int_{Q D}
\frac{1}{\left|\mathbf{r}-\mathbf{r}^{\prime}\right|}
\hat{\mathbf{n}}_{3} \cdot d \mathbf{S}^{\prime}
\end{eqnarray}
with
\begin{equation*}
J=\frac{-\epsilon_{0} A\left(2 e_{15}-e_{33}+e_{31}\right)}{8 \pi
    \varepsilon_{r} \varepsilon_{0}}
\end{equation*}
and
\begin{equation*}
K=\frac{\epsilon_{0}}{8 \pi \varepsilon_{0} \varepsilon_{r}}\left[4
e_{31}+2 e_{33}-A\left(2 e_{15}+e_{31}+e_{33}\right)\right]\,\, .
\end{equation*}
The piezoelectric coefficients are denoted by $e_{ij}$. The
potential due to spontaneous polarization is obtained from
\begin{equation*}
\varphi_{s p o}(\mathbf{r})=\frac{1}{4 \pi \varepsilon_{r}
    \varepsilon_{0}}\left[\int_{Q D} \frac{\mathbf{P}_{Q D} \cdot d
    \mathbf{S}^{\prime}}{\left|\mathbf{r}-\mathbf{r}^{\prime}\right|}+\int_{M}
\frac{\mathbf{P}_{M} \cdot d
    \mathbf{S}_{M}}{\left|\mathbf{r}-\mathbf{r}_{M}\right|}\right]\,\, .
\end{equation*}
The dielectric constant $\epsilon_r$ and $\mathbf{P}_{Q D}$
($\mathbf{P}_{M}$) is the constant spontaneous polarization of the
dot (barrier) material. To obtain the different relevant material
parameters we use a linear interpolation between values of binary
materials InN and GaN weighted with the In/Ga content.

We stress again, that such a VCA description with an underlying WZ
lattice, using dot geometries of high symmetry (e.g. truncated cones
or lens-shaped geometries) and strain and polarization fields from a
surface integral approach will result in an ideal $C_{3v}$ symmetric
system. Thus, the results from the introduced VCA model presents an
ideal reference point to study any symmetry breaking effect of alloy
fluctuations treated in the fully atomistic model discussed above.
Again, the purpose of the VCA model is \emph{not} to match
transition energies; preserving symmetries is the main purpose of
our VCA studies.

\section{Many-body calculations} \label{sec:Many_Body_calculations}

The theoretical analysis of optical properties of semiconductor QDs
is inherently a many body problem. Here, we apply the configuration
interaction (CI) scheme to gain insight into the optical properties
of $c$-plane InGaN/GaN QDs. In the CI method the many-body
Hamiltonian is constructed in the basis of anti-symmetrized products
of (bound) single-particle electron and hole states. In general, the
many-body Hamiltonian $H_\text{MB}$ describing a system $N_e$
electrons and $N_h$ holes is given by:
\begin{eqnarray}
\nonumber
H_\text{MB}&=& \sum_{i}E^{i}_ec^{\dagger}_ic_i+\sum_{\alpha}E^{\alpha}_hh^{\dagger}_\alpha h_\alpha\\
\nonumber
& & +\frac{1}{2}\sum_{ijkl}V^{ee}_{ijkl}c^{\dagger}_ic^{\dagger}_jc_kc_l+\frac{1}{2}\sum_{\alpha\beta\gamma\delta}V^{hh}_{\alpha\beta\gamma\delta}h^{\dagger}_\alpha h^{\dagger}_\beta h_\gamma h_\delta\\
& & -\sum_{il}\sum_{\beta\gamma}\left(V^{eh,di}_{i\beta\gamma
    l}-V^{eh,ex}_{i\beta l\gamma}\right)c^{\dagger}_ih^{\dagger}_\alpha
h_\gamma c_l\,\, . \label{eq:MB_Hamiltonian}
\end{eqnarray}
Here, the operator $c^{\dagger}_i$ ($c_i$) describes/denotes the
creation (annihilation) of an electron in the single-particle state
$i$; $h^{\dagger}_\alpha$ ($h_\alpha$) create (annihilate) a hole in
the state $\alpha$. The two terms in the first line of
Eq.~(\ref{eq:MB_Hamiltonian}) accounts for the electron, $E^{i}_e$,
and hole, $E^{i}_h$, single-particle energies. The terms in line 2
of Eq.~(\ref{eq:MB_Hamiltonian}) describe electron-electron and
hole-hole repulsive Coulomb interactions; we note that since we are
dealing with excitons ($N_e=1$, $N_h=1$) in this work, these terms
are not relevant here. The last line in Eq.~\ref{eq:MB_Hamiltonian}
describes both the direct ($di$) and exchange ($ex$) electron-hole
interaction. The single-particle energies $E^{i}_e$ and
$E^{\alpha}_h$ are obtained from our TB calculations. The Coulomb
matrix elements $V^{eh,di}$ and $V^{eh,ex}$ are directly calculated
from the TB wave functions, as detailed in
Refs.~\cite{ScSc2006,Patra2020}.

In the discussions presented in Sec.~\ref{sec:SPstates_energies}, we
pointed out that hole states are strongly affected in the presence
of alloy fluctuations and the energy level splitting of hole states
is smaller as compared to electrons. Additionally, it was noted that
the energetic difference between hole states also varies between
configurations. Therefore, it is not immediately clear how different
configurations will be affected by randomness in the alloy. If
energetic difference between the hole states are small, one could
expect that more hole states will have to be taken into account in
the CI expansion to obtain accurate results for the FSS values.
Keeping this in mind, we include more hole states (first ten states)
in our CI framework. For the electrons we include the six
energetically lowest single-particle states. We stress that we are
mainly interested in how the alloy microstructure affects the
symmetry and degeneracies of the excitonic states. For this purpose
the above outlined approach is sufficient.

Finally, once the many-body Hamiltonian is constructed and
diagonalized, the resulting many-body states are used to calculate
optical spectra. This is done via Fermi's Golden Rule; the required
dipole matrix elements are again directly obtained from the TB wave
functions. In this work we have restricted the dipole matrix element
calculation to the envelope part, since again we are mainly
interested in symmetry properties of the allowed transitions and not
the absolute magnitude. For such a purpose the envelope part is
sufficient. More details about the calculations of the dipole matrix
elements and emission/absorption spectra can be found
elsewhere.~\cite{ScSc2006,ScSc2008}



\providecommand{\newblock}{}

\end{document}